\documentclass [twocolumn,aps,prb,showpacs]{revtex4}
\usepackage{epsfig,graphicx,amssymb,color,amssymb,amsmath,bm}

\newcommand{\lc}{\lowercase}

\begin{document}
\title{Numerical analysis of carrier multiplication mechanisms in nanocrystal and bulk forms of PbSe and PbS}
\author{Kirill A. Velizhanin}\email{kirill@lanl.gov}
\author{Andrei Piryatinski}\email{apiryat@lanl.gov}

\address{Center for Nonlinear Studies (CNLS),
 Theoretical Division, Los Alamos National Laboratory, Los Alamos, NM 87545 }

\date{\today}

\begin{abstract}
We report on systematic numerical study of carrier multiplication (CM) processes in spherically symmetric nanocrystal (NC) and bulk forms of PbSe and PbS representing the test bed for understanding basic aspects of CM dynamics. The adopted numerical method integrates our previously developed Interband Exciton Scattering Model (IESM) and the effective mass based electronic structure model for the lead chalocogenide semiconductors. The analysis of the IESM predicted CM pathways shows complete lack of the pathway interference during the biexciton photogeneration. This allows us to interpret the biexciton photogeneration as a single impact ionization (II) event and to explain major contribution of the multiple II events during the phonon-induced population relaxation into the total quantum efficiency (QE). We investigate the role of quantum confinement on QE, and find that the reduction in the biexciton density of states (DOS) overruns weak enhancement of the Coulomb interactions leading to lower QE values in NCs as compared to the bulk on the absolute photon energy scale. However, represented on the photon energy scale normalized by corresponding band gap energies the trend in QE is opposite demonstrating the advantage of NCs for photovoltaic applications. Comparison to experiment allows us to interpret the observed features and to validate the applicability range of our model. Modeling of QE as a function of pulse duration shows weak dependence for the gaussian pulses. Finally, comparison of the key quantities determining QE in PbSe and PbS demonstrates the enhancement of II rate in the latter materials. However, the fast phonon-induced carrier relaxation processes in PbS can lead to the experimentally observed reduction in QE in NCs as compared to PbSe NCs.  
\end{abstract}

\maketitle


\section{Introduction}
\label{intro}

An ability to produce more then one electron-hole pairs (excitons) per single absorbed photon in semiconductor materials can potentially enhance the solar energy conversion efficiency in photovoltaic devices beyond the fundamental Shockley-Queisser limit.\cite{kolodinski93,nozik02,schaller04,hanna06,klimov06,beard08a,nozik08,luther08} In the literature this process is referred to as Carrier Multiplication (CM)  or  multiple exciton generation, and has been extensively studied in the {\em bulk} semiconductors for decades.\cite{koc57,smith58,vavilov59,tauc59,cristensen76} Recent ultrafast optical studies of CM in semiconductor {\em nanocrystals} (NC) reported significantly higher QE and much lower activation energy threshold (AET) as compared to the bulk materials on the photon energy scale normalized by corresponding band gap values.\cite{schaller04,ellingson05,schaller05,schaller05a,schaller06,schaller06a,schaller06b,murphy06,schaller07,beard07,pijpers07}   

These reports stimulated extensive experimental and theoretical studies of the CM in the nanostructured materials in which the exciton dynamics is affected by the quantum confinement. It was initially expected that  the quantum confinement enhancement of the Coulomb interactions between the carriers could result in more efficient multi-exciton production. Furthermore, a relaxation of the quasi-momentum conservation constrains by breaking the translational symmetry should open additional pathways for CM and reduce the AET. Finally, the presence of the phonon bottleneck in NCs should slow down the intraband phonon-assisted cooling and further increase the QE. However, the subsequent reports claimed significantly lower QE and even the absence of the CM in NCs.\cite{nair07,nair08,benlulu08,pijpers08} This controversy possibly rises from experimental inaccuracy,\cite{benlulu08,franceschetti08} sample-to-sample variation in surface preparation,\cite{beard09,kilina09a,allan09,tyagi11,nootz11} and contribution from extraneous effects such as photocharging.\cite{mcguire08,mcguire10} The optical measurements of QE in bulk PbS and PbSe semiconductors\cite{pijpers09} show that, in fact, the {\it bulk} QE exceeds validated values in NCs\cite{trinh08,ji09,nair08,mcguire08,mcguire10,trinh11} if compared on the absolute photon energy scale.~\cite{pijpers09} However, due to confinement induced increase in the band gap energy, $E_g$, the benefits to photovoltaics is higher in NCs than in bulk. The controversy calls for development of new sensitive spectroscopic tools,\cite{roslyak10,roslyak10a} alternative photo-current measurements,\cite{sambur10,semonin11} and the reassessment of the quantum confinement role in the CM dynamics based on a unified theoretical model.\cite{nair08,nair11}  

Accepted theoretical models of CM are based on the many-body Coulomb interactions which correspond to the valence-conduction band transitions conserving the total charge but not the number of electrons and holes.\cite{dai06,rabani10,piryatinski10} In the exciton picture, this represents the transitions between the exciton bands of different multiplicity. Hence, we will refer to these transitions below as the {\em interband exciton transitions}, and to the corresponding Coulomb interactions as the {\em interband Coulomb interactions}. Within adopted nomenclature, the {\em intraband Coulomb interactions} (i.e., restricted to an exciton band of a certain multiplicity) conserve the total number of electrons and holes and determine the multi-exciton interaction (e.g., binding) energy.      

The simplest theoretical model initially developed to explain CM in bulk materials is the impact ionization (II) model.\cite{antoncik78} It treats CM dynamics following high energy photoexcitation as the lowest order interband transitions whose rate is given by the Fermi's Golden Rule. This rate depends on the interband Coulomb matrix elements and the final multiple-carriers (e.g., biexciton) density of states (DOS).\cite{kane67,landsberg03} The model has been further applied to interpret the CM dynamics in NCs along with various methods for the electronic structure calculations.   

Franceschetti, An and Zunger performed atomistic pseudopotential calculations of the II and Auger recombination rates in a PbSe NC. Assuming the absence of the quasi-momentum constraint and rapid growth of the biexciton DOS, they estimated low value ($\sim2.2E_g$) of AET defining this quantity as the energy above which the II rate exceeds the Auger recombination one.\cite{franceschetti06} Rabani and Baer performed screened semiempirical pseudopotential calculations of the II rates in CdSe and InAs and found significant reduction of the QE with the NCs size increase. Observed trend makes CM already inefficient in NCs of diameter $\sim 3$~nm. The observation has been rationalized by strong size dependence of the interband Coulomb interactions and the trion DOS behavior.\cite{rabani08}

Using the semiempirical tight binding model Allan and Delerue performed extensive calculations for PbSe and PbSe, InAs and Si NCs.\cite{allan06,allan08} They found that although the Coulomb interactions are enhanced by the quantum confinement, the quantum-confinement-induced reduction in the biexciton DOS facilitates the {\em reduction} of the II rate and subsequently of QE. The comparison of the II rates calculated in PbSe bulk and PbSe NCs shows that the II rates in the NCs does not exceed the bulk one.\cite{allan06} Based on these calculations, the authors of Ref.~[\onlinecite{pijpers09}] argue that the experimentally observed drop of QE in NCs can be attributed to the dominant effect of the reduced carrier's DOS. Lack of the phonon bottleneck, leads to the rapid intraband phonon-assisted relaxation which further reduces the QE.\cite{allan06,allan08}

Schaller, Agranovich, and Klimov, first pointed out that in addition to II, the direct photogeneration of biexcitons can take place in NCs through ``virtual-exciton channel".\cite{schaller05} Subsequently, Rupasov and Klimov suggested that additional contribution to photogeneration QE can rise from the ``vitual biexciton channel".\cite{rupasov07} Using the effective mass model for the electronic structure of PbS and PbSe NCs due to Kang and Wise,\cite{kang97} the authors estimated the photogeneration QE and argued that this process dominates the CM dynamics in NCs. Recently, Silvestri and Agranovich using the same model, performed detailed calculations for relatively small radii (specifically for 1.95~nm and 3~nm) PbSe NCs.\cite{silvestri10}  They concluded that the contribution of the photogeneration processes in NCs is much weaker than it was claimed before, and clarified that the overestimated QE results from disregarding the effect of selection rules for the interband Coulomb matrix elements and the oscillator strength factors weighting optically allowed transitions.  Here, we demonstrate that the absence of the interference between the photogeneration pathways and the presence of small size dispersion further reduces the photogeneration QE. 
 
Another model describing the {\it coherent} photogeneration of biexcitons from the resonant exciton states initially proposed by Shabaev, Efros, and Nozik considered idealized five level system,\cite{shabaev06} and was subsequently refined to account for the biexciton DOS effect.\cite{witzel10} Using the effective mass ${\bf k\cdot p}$ Hamiltonian, the authors of Ref.~[\onlinecite{witzel10}] performed calculations for small radii (i.e., 2~nm and 3~nm) PbSe NCs, and found that dense biexciton DOS leads to the vanishing coherent oscillations. They also pointed out that efficient CM can be expected in considered small NCs. The drawback of the coherent superposition model is the lack of the pure-dephasing effects and the inhomogeneous broadening which as  we demonstrate here play important role in the CM dynamics reducing the QE. 

Finally, the {\it ab initio} calculations on small {\it clusters} ($\lesssim 1$~nm)  demonstrated the role of strong Coulomb correlations and fast exciton-biexciton dephasing in multiexciton photogeneration,~\cite{prezhdo09,fischer10} and suggested contributions of the phonon-assisted Auger processes to CM.\cite{hyeon11} The atomistic calculations mostly focused on II dynamics occurring during the population relaxation are limited to the small diameter ($\lesssim 3$~nm) NCs. However, reported experimental studies consider larger NCs with the diameter varying up to 10~nm. To close this gap and to perform comparison with bulk, an extrapolation procedure combined with atomistic calculations has been used.\cite{delerue10}  The photogeneration dynamics has been investigated using effective mass models for larger but still small diameter ($\sim 4-6$~nm) NCs. 

Recently, we have published a letter summarizing the results of our numerical investigation on the direct photogeneration and population relaxation processes contributing to CM in a broad diameter range PbSe NCs and in PbSe bulk.\cite{velizhanin11a} The calculations employ our earlier proposed Interband Exciton Scattering Model (IESM)\cite{piryatinski10} which recovers the models discussed above as limiting cases and which we further parameterized using the effective mass, ${\bf k\cdot p}$, electronic structure model proposed by Kang and Wise.\cite{kang97}$^{,}$\footnote{The IESM we use is a complimentary approach to the Hilbert space Green's function method independently proposed by Rabani and Baer.\cite{rabani10} In contrast to our calculations, the latter method has been used along with the atomistic electronic structure model and subsequently applied to the smaller size NCs.} In the letter, we argue that in both cases of the photogeneration and population relaxation the II is the main mechanisms of CM. This explains weak contribution of the direct photogeneration to the total quantum QE.  Analyzing the scaling of the total QE with the NC size, we found that QE in NCs plotted on absolute energy scale does not exceed that in the bulk. This is in agreement with reported experimental data and some theoretical studies, and confirms that the quantum-confinement-induced reduction in the biexciton DOS makes a dominant contribution
to QE. 

Present paper discusses in great details the theoretical method and provides extended analysis of the numerical results summarized in Ref.~[\onlinecite{velizhanin11a}]. In addition, we consider the effect of the optical pump pulse duration on QE, and provide a comparative analysis of CM in PbS and PbSe materials. The paper is organized as follows. In Sec.~\ref{model}, we review the weak Coulomb interaction limit of the IESM used for numerical calculations, introduce convenient quasicontinuous representation, and further discuss the details of the model parameterization. In Sec.~\ref{numerics} an extensive analysis of our numerical results on PbSe NCs is given. Comparison of the key CM features in PbSe and PbS is performed in Sec.~\ref{comp}. Sec.~\ref{disc} concludes the paper by a discussion on the limitations of the adopted model, connection with experiment, and possible improvements of QE in NCs.


\section{Theoretical Model and Numerical Parameterization}
\label{model}

The central quantity describing CM efficiency is QE which can be calculated as
\begin{eqnarray}
\label{QE}
    QE =\frac{2N_{xx}+N_x}{N_{xx}+N_x}.
\end{eqnarray}
Here $ N_x$ and $N_{xx}$ are the total exciton and biexciton populations, respectively, produced by single absorbed photon in the limit of vanishing pump fluencies. They depend on the delay time measured from the center of the pump pulse and allow one to determine both the QE due to the photogeneration event and the total QE after the population relaxation. A complimentary quantity, often used to describe CM yield, is the biexciton quantum yield, $\eta_{xx}=QE-1$, defined as a number of biexcitons produced per single absorbed photon.

In this section, we describe the weak Coulomb limit of the IESM\cite{piryatinski10} which we use to calculate the time evolution of the exciton and biexciton populations entering Eq.~(\ref{QE}). We also introduce the quasicontinuous frequency representation which is convenient to analyze the quantum confinement signatures and the size scaling of the quantities determining the QE in transition from NC to the bulk limit. Finally, we discuss the model parameterization for the numerical calculations of QE in PbSe and PbS materials.

\subsection{Biexciton Photogeneration and Population Relaxation Model}
\label{weakc}

\begin{figure*}[t]
	\centering
	\includegraphics[width=6.0in,clip]{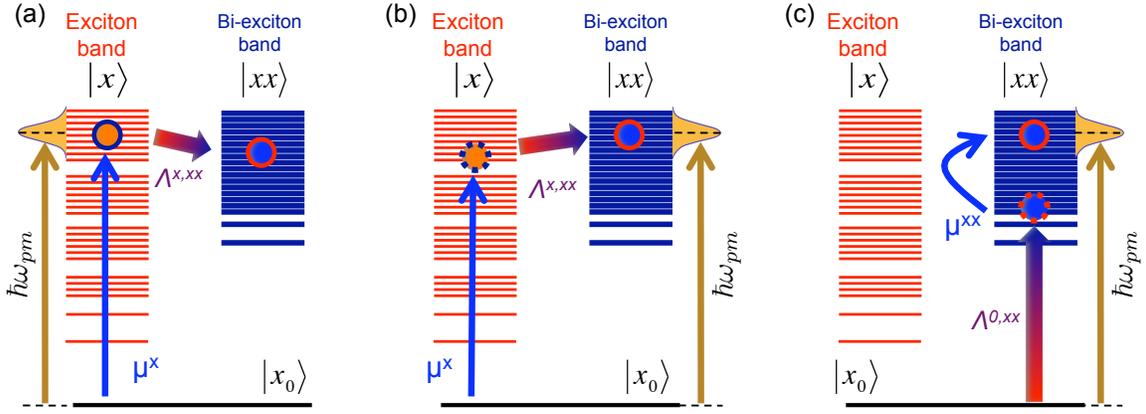}
	\caption{Non-interfering biexciton photogeneration pathways associated with the weak Coulomb limit. (a) Indirect biexciton 				photogeneration via exciton states. (b) Direct biexciton photogeneration via virtual exciton states. (c) Direct 			biexciton photogeneration via biexciton states. (See details in the text.)}
	\label{Fig-spath}
\end{figure*}	

In the weak Coulomb interaction limit, the leading contribution to the population of the $a$-th exciton state generated by the pump pulse is given by
\begin{eqnarray}
\label{nx}
    {n^{x}_a}^{(0)}&=&|\mu^x_{a0}|^2 {\cal I}(\omega^x_a-\omega_{pm}),
\end{eqnarray}
where $\mu^x_{a0}$ is the transition dipole moment between the ground state, $|x_0\rangle$, (i.e., filled valence band) and an exciton state $|x_a\rangle$, and
\begin{eqnarray}
\label{Ip}
    {\cal I}(\omega^x_a&-&\omega_{pm})=
    \frac{2}{\hbar^2}\int_{-\infty}^\infty dt^{'}\int_{0}^\infty dt_1
    e^{-\gamma^x_{a}t_1}
    \\\nonumber&\times&
    \cos\left[\left(\omega^x_a-\omega_{pm}\right)t_1\right]
   {\cal E}_{pm}(t^{'}) {\cal E}_{pm}(t^{'}-t_1),
   \end{eqnarray}
is the pulse self-convolution function depending on the ground state to exciton state transition frequency, $\omega^x_a$, and dephasing  rate, $\gamma^x_a$. The pump pulse is characterized by the envelope amplitude, ${\cal E}_{pm}(t)$, giving the temporal profile, and the central frequency, $\omega_{pm}$. In our calculations, we use the Gaussian form of the pulse envelope function, ${\cal E}_{pm}(t) = {\cal E}^{(0)}_{pm} e^{-t^2/2\tau^2_{pm}}$, with the mean amplitude, ${\cal E}^{(0)}_{pm}$, and pulse duration, $\tau_{pm}$. For the pulses longer than typical dephasing time, $\tau_{pm}\gg1/\gamma$, the continuous wave (cw) limit is recovered and the pulse self-convolution function becomes proportional to the Lorentzian line shape function.

To be consistent with the calculations of the photogenerated {\em biexciton} population appearing in the second order interband Coulomb coupling, we account for the first and second order corrections to the {\em exciton} population and disregard the effect of the exciton coherences both introduced in Ref.~[\onlinecite{piryatinski10}]. The negligible contribution of the latter quantities has been validated by our direct numerical calculations.

The photogenerated population of the $k$-th biexciton state is\cite{piryatinski10}
\begin{eqnarray}\label{nxx}
   &~&{n^{xx}_k}^{(2)}= \sum_{a\geq1}
                \left|\Lambda^{xx,x}_{k,a}\mu^x_{a0}\right|^2{\cal I}(\omega^x_a-\omega_{pm})
 \\\nonumber &+&
    \left|\sum_{a\geq1}\Lambda^{xx,x}_{k,a}\mu^x_{a0}
    +\sum_{l\geq1}\mu^{xx}_{kl}\Lambda^{xx,x}_{l,0}
    \right|^2
    {\cal I}(\omega^{xx}_k-\omega_{pm}),
\end{eqnarray}
where $\mu^{xx}_{kl}$ is the intraband transition dipole between the biexciton states $|xx_k\rangle$ and $|xx_l\rangle$, and
\begin{eqnarray}
  \label{Lmbd}
    \Lambda^{xx,x}_{k,a}&=& \frac{V^{xx,x}_{k,a}}
            {\hbar\left[\omega^{xx}_k -\omega^x_a+ i\gamma^{xx,x}_{k,a}\right]},
\end{eqnarray}
describes the transition amplitude associated with the single (Born) scattering event between exciton sate, $|x_a\rangle$, and biexciton state, $|xx_k\rangle$. The quantity $\Lambda^{xx,x}_{l,0}$, also entering Eq.~(\ref{nxx}), is the transition amplitude between the ground state, $|x_0\rangle$ ($\omega^x_0=0$), and biexciton state, $|xx_l\rangle$. $V^{xx,x}_{k,a}$ is the interband Coulomb matrix element in the exciton representation giving rise to the interband scattering, and $\gamma^{xx,x}_{k,a}$ is the interband dephasing rate. The pulse self-convolution function, ${\cal I}(\omega^{xx}_k-\omega_{pm})$, entering Eq.~(\ref{nxx}) contains $\omega^{xx}_k$ and $\gamma^{xx}_k$ instead of $\omega^{x}_k$ and $\gamma^{x}_k$, respectively. According to our calculations the coherence contributions to the biexciton populations\cite{piryatinski10} are also negligible.

Our numerical calculations show that the rapid sign-variation of the interfering terms in Eq.~(\ref{nxx}) leads to their cancelation allowing us to write Eq.~(\ref{nxx}) as
\begin{eqnarray}\label{nxxi}
    {n^{xx}_k}^{(2)}&=& \sum_{a\geq1}
                \left|\Lambda^{xx,x}_{k,a}\mu^x_{a0}\right|^2{\cal I}(\omega^x_a-\omega_{pm})
 \\\nonumber &+&
    \sum_{a\geq1}\left|\Lambda^{xx,x}_{k,a}\mu^x_{a0}\right|^2{\cal I}(\omega^{xx}_k-\omega_{pm})
 \\\nonumber &+&
    \sum_{l\geq1}\left|\mu^{xx}_{kl}\Lambda^{xx,x}_{l,0}
    \right|^2{\cal I}(\omega^{xx}_k-\omega_{pm}).
  \end{eqnarray}
The resulting three {\em non-interfering} photogeneration pathways are illustrated in Fig.~\ref{Fig-spath}.

The first and the second pathways, given by the first and the second terms in Eq.~(\ref{nxxi}), are shown in panels~(a) and (b), respectively. These terms describe redistribution of the exciton oscillator strengths ($\sim|\mu_{0a}^{x}|^2$) between exciton and biexciton bands mediated by the interband Born scattering ($\Lambda^{x,xx}$). First pathway (panel~(a)), involves the ground-state-to-exciton resonant optical transition and further scattering to final biexciton states distributed around $\hbar\omega_{pm}$ according to the non-zero components of $\Lambda^{x,xx}$. We will refer to this process as the {\em indirect biexciton photogeneration} throughout this paper.\cite{piryatinski10} Panel~(b) illustrates the second pathway, where the exciton is virtual and final biexciton state is in resonance with the optical pulse. Thus, this pathway will be refereed to as the {\it direct biexciton photogeneration via virtual exciton states}.\cite{schaller05} The third pathway (panel~(c)) consists of Born scattering between the ground state and biexciton states ($\Lambda^{0,xx}$) stabilized by the intraband dipole transition, $\mu^{xx}$. Accordingly, the final biexciton state is in resonance with the optical pulse. This process, refereed to as the {\it direct biexciton photogeneration via biexciton states},\cite{rupasov07} becomes prohibited in the bulk limit by the momentum conservation constraint for optical valence-conduction band transitions.

Equation~(\ref{nx}) with the higher order corrections and Eqs.~(\ref{nxx})--(\ref{nxxi}) fully describe the exciton and biexciton populations prepared by the pump pulse. We use them for the numerical evaluation of the photogenerated exciton, $N_x=\sum_{a}n^{x}_a$, and biexciton, $N_{xx}=\sum_kn^{xx}_k$, populations and QE (Eq.~(\ref{QE})). They also provide the initial conditions to model the population relaxation dynamics using a set of rate equations.\cite{piryatinski10} Using this computational approach, our goal is to clarify the effect of quantum confinement on QE in transition from NCs to the bulk limit. Since the CM dynamics in NCs takes place in the energy region characterized by high electron and hole DOS, we expect that the CM dynamics in NCs and in the bulk should have common features. Thus, we intend to see how strongly these features are affected by the presence of the confinement potential.       

Quantitatively, we are going to look at the interplay between the size scalings of the interband Coulomb interaction and exciton/biexciton DOS determining the QE variation in transition from NC to the bulk limit. First, we define the bulk limit as the thermodynamic limit: $V\rightarrow\infty$, $V/v\rightarrow\infty$, and $v=const.$, where $V$ is a crystal volume, $v$ is the unit cell volume, and the ratio $V/v$ gives the number of unit cells in the crystal. Then Eqs.~(\ref{nx}) and (\ref{nxxi}), and the quantities determining the population relaxation should be represented in such a form that the effect of the interband Coulomb interactions and exciton/biexciton DOS are clearly distinguished. This can be achieved by using the quasicontinuous energy representation. Since some quantities of interest in the bulk limit have volume scaling, it is convenient to introduce associated intensive (i.e., volume-independent in the bulk limit) variables. Their deviation from the well defined bulk values will provide us with the convenient measure of the quantum confinement effects. 

We start with the exciton and biexciton populations which in the quasicontinuous energy representation read
\begin{eqnarray}
\label{nxE}
n_x(\omega)&=&\sum_{a\geq1} n^{x}_a\delta(\omega-\omega^{x}_a),
\\\label{nxxE}
n_{xx}(\omega)&=&\sum_{k\geq1} n^{xx}_k\delta(\omega-\omega^{xx}_k),
\end{eqnarray}
respectively. Both of them have linear scaling with the volume in the bulk limit. Therefore, we eliminate the latter scaling by multiplying the total populations with the dimensionless prefactors ($v/V$) and end up with the following intensive quantities
\begin{eqnarray}
\label{Nx}
\tilde N_x &=& \frac{v}{V}\int^\infty_0 n_x(\omega) d\omega,
\\\label{Nxx}
\tilde N_{xx}&=&\frac{v}{V}\int^\infty_0 n_{xx}(\omega) d\omega.
\end{eqnarray}
Although the prefactor $v/V$ cancels out in the expression for QE (Eq.~(\ref{QE})), indicating that the latter quantity is indeed intensive, it is convenient to keep it in Eqs.~(\ref{Nx}) and (\ref{Nxx}) for consistency.

According to Appendix~\ref{Appx_vscl}, we define the intensive exciton and biexciton DOS as
\begin{eqnarray}\label{XDOS}
	\rho_{x}(\omega)&=&\left(\frac{v}{V}\right)^2\sum_a\delta(\omega-\omega^x_a),
\\\label{XXDOS}
	\rho_{xx}(\omega)&=&\left(\frac{v}{V}\right)^4\sum_k\delta(\omega-\omega^{xx}_k),
\end{eqnarray}
respectively. The associated {\em optically allowed} exciton and joint biexciton DOS read
\begin{eqnarray}\label{XODOS}
	\tilde\rho_{x}(\omega)&=&\left(\frac{v}{V}\right)\sum_a|\mu^x_{a0}|^2\delta(\omega-\omega^x_a),
\\\label{XXODOS}
	\tilde\rho_{xx}(\omega_1,\omega_2)&=&\left(\frac{v}{V}\right)^4
		\sum_{kl}|\mu^{xx}_{kl}|^2
\\\nonumber&\times&		
		\delta(\omega_1-\omega^{xx}_k)\delta(\omega_2-\omega^{xx}_l),
\end{eqnarray}
respectively. They carry information on the optical selection rules reducing the number of states participating in the photogeneration process.

Our central quantity is the effective Coulomb term defined as the r.m.s. of the interband Coulomb matrix elements which couple states within the frequency intervals $[\omega_1,\omega_1+d\omega_1]$ and $[\omega_2,\omega_2+d\omega_2]$,
\begin{eqnarray}\label{Veff}
	V^{x,xx}_{eff}(\omega_1,\omega_2)=\left(\frac{V}{v}\right)^2\left[\sum_{a,m}\left|V^{x,xx}_{a,m}\right|^2
\right.&~&\\\nonumber\left.\times
	\frac{\delta(\omega_1-\omega^x_a)\delta(\omega_2-\omega^{xx}_m)}
			{\sum_{b}\delta(\omega_1-\omega^x_b)\sum_{n}\delta(\omega_2-\omega^{xx}_n)}\right]^{1/2}.
\end{eqnarray}
Related effective Coulomb term, coupling the ground state and biexciton states, can also be defined as
\begin{eqnarray}\label{Veff0}
	V^{xx}_{eff}(\omega)=\left(\frac{V}{v}\right)^{3/2}\left[\sum_{m}\left|V^{x,xx}_{0,m}\right|^2
	\frac{\delta(\omega-\omega^{xx}_m)}{\sum_{n}\delta(\omega-\omega^{xx}_n)}\right]^{1/2}.
\end{eqnarray}

As we discuss in Appendix~\ref{Appx_vscl}, the volume prefactor $(V/v)^2$ [$(V/v)^{3/2}$] in Eq.~(\ref{Veff}) [Eq.~(\ref{Veff0})]  corresponds to a finite effective Coulomb value in the bulk limit. Therefore, the size scaling of such defined interaction with the NC diameter, $d$, provides quantitative measure of the quantum confinement effects. In general, a deviation from the bulk value for the effective Coulomb reflects the net result of the scaling of the Coulomb matrix elements with $d$, relaxation of the momentum conservation constraints, and the appearance of new selection rules associated with symmetry of the confinement potential.  

Assuming the cw excitation, the exciton population produced by the pump pulse is simply proportional to the corresponding optically allowed DOS (Eq.~(\ref{XODOS})),
\begin{eqnarray}\label{NxE}
\tilde N_{x}(\omega_{pm})={\cal A}\tilde\rho_x(\omega_{pm}),
\end{eqnarray}
where ${\cal A}=2\pi^{3/2}\tau_{pm}{{\cal E}^{(0)}_{pm}}^2/\hbar^2$. The biexciton population as a function of the pump frequency in the adopted representation reads
\begin{eqnarray}\label{NxxE}
&~&\tilde N_{xx}(\omega_{pm})=
\\\nonumber&~&
\frac{{\cal A}}{\hbar^2}\int d\omega^{'}[V^{x,xx}_{eff}(\omega_{pm},\omega^{'})]^2
	\frac{\tilde\rho_{x}(\omega_{pm})\rho_{xx}(\omega^{'})}{(\omega^{'}-\omega_{pm})^2+\gamma^2}
\\\nonumber&+&	
	\frac{{\cal A}}{\hbar^2}\int d\omega^{'}[V^{x,xx}_{eff}(\omega^{'},\omega_{pm})]^2
	\frac{\tilde\rho_{x}(\omega^{'})\rho_{xx}(\omega_{pm})}{(\omega^{'}-\omega_{pm})^2+\gamma^2}
\\\nonumber&+&	
\frac{{\cal A}}{\hbar^2}\int d\omega^{'}[V^{xx}_{eff}(\omega^{'})]^2
	\frac{\tilde\rho_{xx}(\omega^{'}\omega_{pm})}{{\omega^{'}}^2}.
\end{eqnarray}
As desired, Eq.~(\ref{NxxE}) is volume independent and clearly distinguish the contributions from the effective Coulomb interactions and variously defined DOS. Accordingly, Eqs.~(\ref{NxE}) and (\ref{NxxE})  provide central expressions for the interpretation of the numerical simulation results on the photogeneration QE as discussed in the subsequent section.

The population relaxation dynamics is described by the following set of kinetic equations
\begin{eqnarray}
\label{nxr}
\dot{n}_x(\omega)&=&-k_{II}(\omega)n_x(\omega)+k_{AR}(\omega)n_{xx}(\omega)
	\\\nonumber
			&-&\int_0^\infty d\omega'\;\Gamma_x(\omega',\omega)n_x(\omega)	\\\nonumber
			&+&\int_0^\infty d\omega'\;\Gamma_x(\omega,\omega')n_x(\omega'),
\\\label{nxxr}
\dot{n}_{xx}(\omega)&=&k_{II}(\omega)n_x(\omega)-k_{AR}(\omega)n_{xx}(\omega)	\\\nonumber
			&-&\int_0^\infty d\omega'\:\Gamma_{xx}(\omega',\omega)n_{xx}(\omega)
	\\\nonumber
			&+&\int_0^\infty d\omega'\: \Gamma_{xx}(\omega,\omega')n_{xx}(\omega').
\end{eqnarray}
These equations contain both the II and Auger recombination rates
\begin{eqnarray}\label{kII}
k_{II}(\omega)&=&\frac{2\pi}{\hbar^2}[V^{x,xx}_{eff}(\omega)]^2\rho_{xx}(\omega),
\\\label{kAR}
k_{AR}(\omega)&=&\frac{2\pi}{\hbar^2}\left(\frac{v}{V}\right)^2[V^{x,xx}_{eff}(\omega)]^2\rho_{x}(\omega),
\end{eqnarray}
respectively. Here the shorthand notation $V^{x,xx}_{eff}(\omega)$ stands for the diagonal component of the effective Coulomb term, i.e., $V^{x,xx}_{eff}(\omega)\equiv V^{x,xx}_{eff}(\omega,\omega)$. According to Eq.~(\ref{kII}), the II rate is an intensive variable. In contrast, the Auger recombination rate (Eq.~(\ref{kAR})) vanishes in the the bulk limit as $d^{-6}$.

In the case where the II processes dominate the CM dynamics, the ratio of the II to Auger recombination rates
\begin{eqnarray}\label{kr}
\frac{k_{II}(\omega)}{k_{AR}(\omega)}=\left(\frac{V}{v}\right)^2\frac{\rho_{xx}(\omega)}{\rho_{x}(\omega)},
\end{eqnarray}
should determine QE. Since this ratio is proportional to the ratio of the corresponding DOS, it has been proposed as a selection criterion for the materials showing efficient CM.\cite{luo08} However, Eq.~(\ref{kr}) clearly shows that besides the material-specific  signatures given by the ratio of the {\em intensive} DOS, it also contains the volume scaling factor. Therefore, we argue that Eq.~(\ref{kr}) can only be used for a material selection criterion, after the volume prefactor is eliminated.

The intraband relaxation rates, $\Gamma^{x}$, $\Gamma^{xx}$, in Eqs.~(\ref{nxr}) and (\ref{nxxr}) describe the phonon-assisted cooling. In the absence of the phonon bottleneck and in the region of high exciton DOS, it is expected that single-phonon processes dominate the exciton and biexciton intraband relaxation. Accordingly, we calculate these quantities by using the following expression\cite{Zwanzig_NSM,grifoni99}
\begin{equation}\label{Gamr}
\Gamma_{i}(\omega',\omega)={\rm sign}(\omega'-\omega)\frac{2J_{i}(|\omega'-\omega|)}{\exp[{\hbar(\omega'-\omega)/k_BT}]-1},
\end{equation}
where $i=x,xx$. $k_BT$ is thermal energy, and $J_{i}(\Delta \omega)$ is the phonon spectral density approximated by the ohmic form with exponential cutoff
\begin{equation}\label{Jph}
J_{i}(\Delta \omega)=\lambda_{i}\frac{\Delta \omega}{\omega_c}e^{-\Delta \omega/\omega_c}.
\end{equation}
Here, the adjustable parameters are electron-phonon coupling, $\lambda_{i}$, with the constraint $\lambda_{xx}=2\lambda_x$ and the phonon frequency cutoff, $\omega_c$.

Eqs.~(\ref{nxr})-(\ref{Jph}) represent the quasicontinuous representation of the discrete kinetic equations introduced in Ref.~[\onlinecite{piryatinski10}]. We found that the former representation is more suitable for numerical integration. The solution of these equations with the initial conditions determined by the photogenerated populations (Eqs.~(\ref{nx})--(\ref{nxxi})) provide closed computational scheme that we use to determine the total QE. Next, we discuss the parameterization of the introduced model and some details on the adopted numerical techniques.

\subsection{Model Parameterization}
\label{mparam}

An accurate knowledge of single electron and hole wave functions is required to construct the exciton and biexciton states and to evaluate transition dipoles and the interband Coulomb matrix elements. For this purpose, we adopt the effective mass ${\bf k\cdot p}$ formalism, originally developed by Mitchell and Wallis\cite{mitchell66} and Dimmock\cite{dimmock71} for the bulk lead chalocogenide semiconductors, and further advanced by Kang and Wise\cite{kang97} for the spherically symmetric NCs. The formalism is based on a four-band envelope function model explicitly taking into account spin-orbit interaction between valence and conduction bands. In considered PbSe and PbS materials, the band structure anisotropy is small, and therefore, neglected in our calculations.

An electron and hole wave function obtained within this formalism reads
\begin{equation}\label{KW_spinor}
\Psi_i({\bf r})=\sum^4_{m=1} F^i_m({\bf r})u_m({\bf r}).
\end{equation}
Here $u_m({\bf r})$ is $m$-th component of the bulk Bloch wave function associated with the band-edge states in $L$-valley whose index, $m=1,\dots,4$, denotes the bands.\cite{mitchell66,dimmock71} The envelope eigenfunctions $F^i_m({\bf r})$ and the corresponding eigenenergies, $\hbar\omega_i$, are found by solving the ${\bf k\cdot p}$ Hamiltonian eigenvalue problem with the infinite wall boundary condition, $F_i(|{\bf r}|=R)=0$, at the surface of a spherical NC of radius $R$.\cite{kang97} If $\hbar\omega_i>0$ ($\hbar\omega_i<0$) we identify the state as a conduction band electron (valance band hole) state. The eigenstate index represents a set of quantum numbers, $i=\{n,\pi,j,m\}$, such as primary quantum number describing number of the wave function nods, parity, total angular momentum, and its projection, respectively. In what follows, we will refer to the single particle states, described by Eq.~(\ref{KW_spinor}), as the Kang-Wise (KW) states. In the bulk limit, we set the envelope functions, $F^i_m({\bf r})$, to the plane waves.

A natural way to introduce the exciton and biexciton states is to use the second quantization representation within the basis of the KW-states (Eq.~(\ref{KW-sq})). Taking into account that the electron-hole Coulomb interactions are weak compared to their kinetic energies (i.e., strong confinement regime), we introduce exciton and bi-exciton states as the following configurations of uncorrelated electron-hole pairs
\begin{eqnarray}\label{x-sq}
|x_a\rangle&=&\hat{c}^\dag_q \hat{d}^\dag_r |x_0\rangle,
\\\label{xx-sq}
|xx_k\rangle&=&\hat{c}^\dag_p \hat{c}^\dag_q \hat{d}^\dag_r \hat{d}^\dag_s |x_0\rangle.
\end{eqnarray}
As introduced above, the ground state, $|x_0\rangle$, is attributed to the fully filled valence band. An exciton (biexciton) energy in terms of the electron, $\hbar\omega^e_p$, and the hole $\hbar\omega^h_q$ energies is simply: $\hbar\omega^x_a = \hbar\omega^e_q-\hbar\omega^h_r$ ($\hbar\omega^{xx}_k =\hbar\omega^e_p+\hbar\omega^e_q-\hbar\omega^h_r-\hbar\omega^h_s$). Note that the exciton index $a=\{p,q\}$ (biexciton index $k=\{pq,rs\}$) is a collection of two (four) sets of KW quantum numbers.

The expressions used to calculate the dipole moments $\mu^x_{a0}$ and $\mu^{xx}_{kl}$, and the interband Coulomb matrix elements, $V^{x,xx},$ in Eqs.~(\ref{nx}), (\ref{nxx}), (\ref{Lmbd}), (\ref{Veff}), and (\ref{Veff0}) using associated matrix elements in the KW basis are given in Appendix~\ref{Appx_mtxl}. Numerical calculations of the interband Coulomb KW matrix elements are performed by using the multipole expansion in terms of the Clebsch-Gordan coefficients. The dephasing rates for the {\em high excited states} are not available from experiment, and in our calculations we use some average value of $\hbar\gamma^x =\hbar\gamma^{xx}/2 =\hbar\gamma^{x,xx}/2 = 50$ meV. This value is consistent with the calculations performed on PbSe clusters.\cite{kamisaka06,kamisaka08,prezhdo09} 

The numerical calculations of the  photogenerated populations have been performed according to Eqs.~(\ref{nx})--(\ref{Lmbd}) including the first and second order corrections to Eq.~(\ref{nx}). To account for the degeneracy of the exciton and biexciton states associated with the four equivalent $L$-valleys, we multiply the r.h.s. of Eq.~(\ref{nx}) and Eq.~(\ref{nxx}) by factors 4 and 16, respectively. This assumes that the inter-valley Coulomb scattering processes are negligible. In most calculations, the pump pulse duration is set to $\tau_{pm}=50$~fs, which is typical value used in the experimental studies.\footnote{The exception is the calculation of the QE dependence on $\tau_{pm}$, where we specifically indicate $\tau_{pm}$ variation range.}

The number of exciton and biexciton states to handle computationally is extremely high. Even for PbSe NCs of moderate size ($d\sim 5$~nm), the number of the  biexciton states with energies below $4E_g$ is $\sim 10^5$ . In the bulk limit, the number of the exciton and and biexciton states becomes infinite rendering the direct summation over these states in Eq.~(\ref{nxx}) impossible. Therefore, in both cases of finite-size NCs and the bulk, we use the Monte Carlo sampling which allows us to consider significantly larger NCs than it was possible in the work of Silvestri and Agranovich.\cite{silvestri10}

To perform numerical calculations of the population relaxation processes, we first evaluated the interband Auger recombination and II rates given by Eqs.~(\ref{kII})--(\ref{kAR}). For this purpose, we smeared the delta-functions in the effective Coulomb term (Eq.~(\ref{Veff})) and exciton/biexciton DOS (Eqs.~(\ref{XDOS}) and (\ref{XXDOS})) with the Lorentzian profiles containing the corresponding dephasing rates and used the Monte Carlo procedure to sum over the exciton and biexciton states. To evaluate the phonon-induced intraband relaxation rates (Eqs.~(\ref{Gamr}) and (\ref{Jph})), we set the room temperature value for $k_BT=25$~meV, the cutoff energy to $\hbar\omega_c=50$~meV. The electron-phonon coupling $\lambda_x=\lambda_{xx}/2$ was fit to reproduce the experimentally observed intraband relaxation time in the range of $0.5\leq\tau_{ph}\leq5$~ps.\cite{allan06,pijpers09} The kinetic equations (Eqs.~(\ref{nxr}) and (\ref{nxxr})) have been further numerically integrated on the energy grid. All calculated observables are averaged over ensemble of NCs with the Gaussian size distribution with the standard deviation $\sigma = 5\%$.


\section{Interplay of CM pathways in P\lowercase {b}S\lowercase{e}}
\label{numerics}

\begin{figure}[t]
\centering
\includegraphics[width=2.95in]{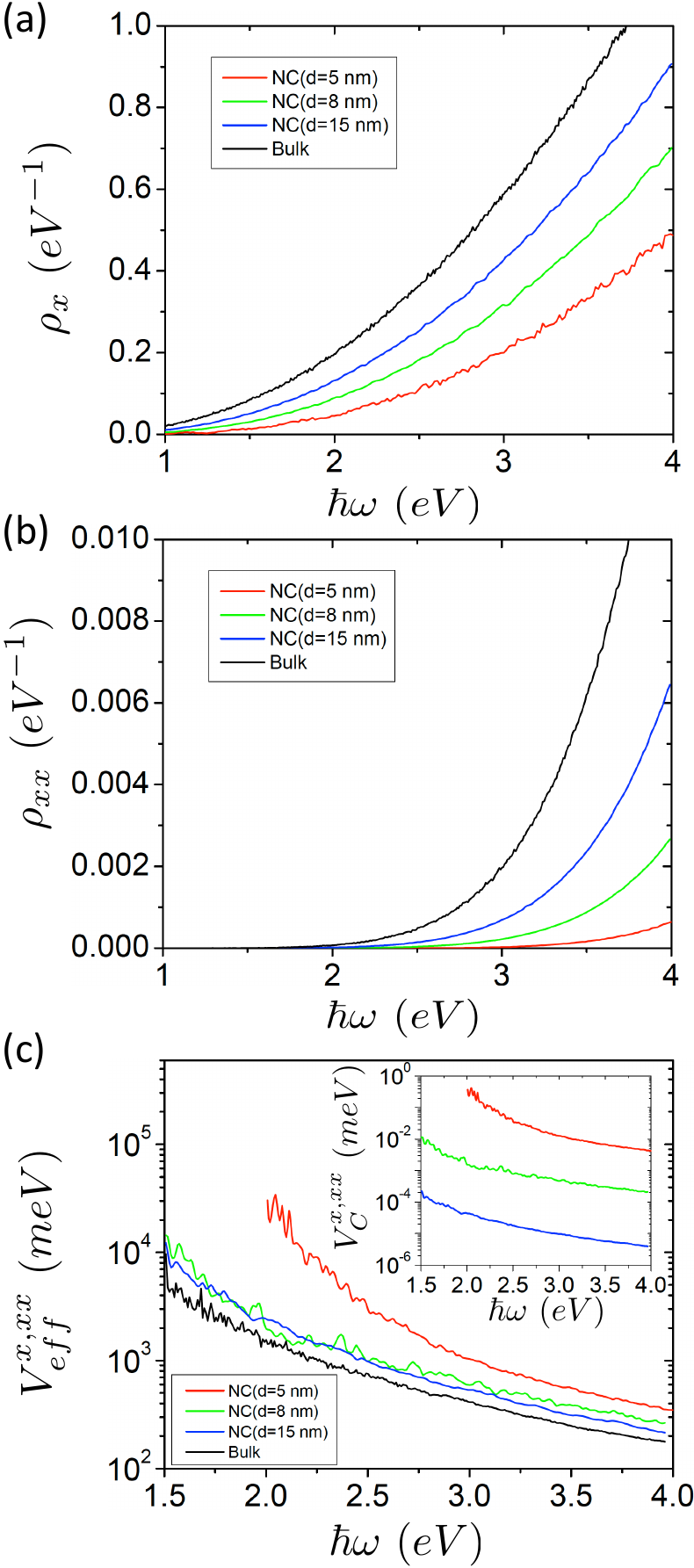}
\caption{Energy dependence of (a) exciton and (b) biexciton DOS calculated for various diameter, $d$, PbSe NCs and the bulk.
			(c) The diagonal component of the effective Coulomb term in PbSe NCs and the bulk. The inset shows actual 
			interband Coulomb interaction in NCs defined as the effective Coulomb term with the volume prefactor added, 
			i.e., $V^{x,xx}_{C}=(V/v)^2V^{x,xx}_{eff}$.}
\label{CDOS}
\end{figure}

In this section, we discuss the results of our numerical calculations of the photogeneration and population relaxation processes in NC and bulk PbSe to clarify the effect of the CM pathways interplay and quantum confinement on QE. As we already mentioned, the QE is determined by the interplay between the DOS and the effective Coulomb scalings. Therefore, to set the stage for our analysis we first consider these quantities separately.

Calculated energy dependence of the exciton and biexciton DOS (Eqs.~(\ref{XDOS})~and~(\ref{XXDOS})) for different diameter PbSe NCs and the bulk limit are shown in panels~(a) and (b) of Fig.~\ref{CDOS}, respectively. Obviously, the quantum confinement leads to the DOS reduction in NCs compared to the bulk. The energy dependence of the DOS follows power law. Specifically for the bulk limit, we find that $\rho_{x}(\omega)\sim (\omega-\omega_g)^{2.2}$ and $\rho_{xx}(\omega)\sim (\omega-2\omega_g)^{6}$ with $\hbar\omega_g=0.28$~eV reflecting contributions of the nonparabolic regions of the electron/hole band structures. Even in the high DOS regions where the bulk behavior is expected in NCs, the DOS converges rather slowly to the limiting bulk values as the NC diameter increases.

\begin{figure}[t]
\begin{center}
\epsfig{file=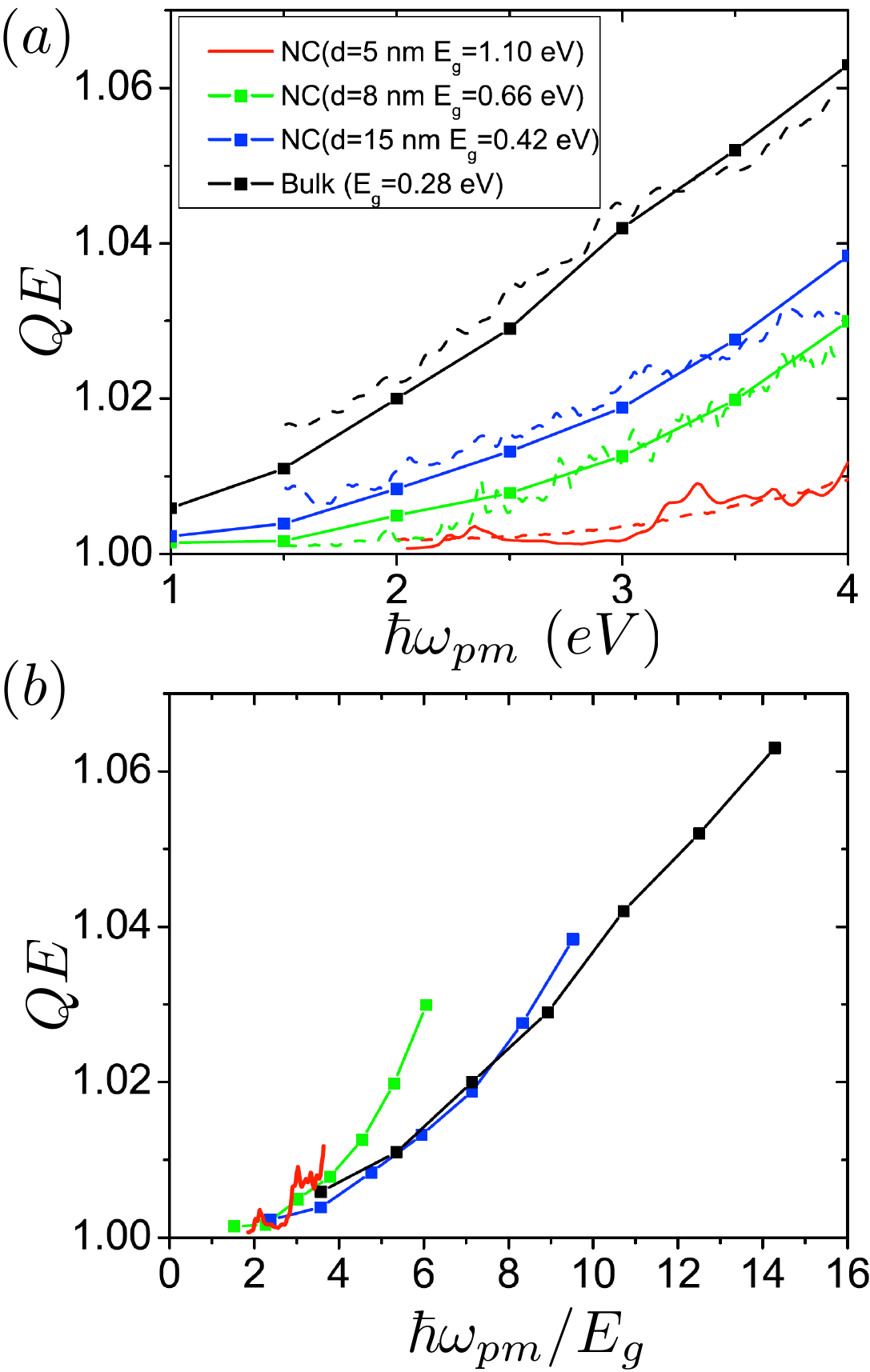,width=3.0in}
\end{center}
\caption{(a) The pump energy dependence of the photogeneration QE for various diameter, $d$, NC and bulk PbSe plotted on the
			absolute pump energy scale. Solid lines 
			represent the results obtained from numerically exact (Eqs.~(\ref{nx})--(\ref{Lmbd})) calculations, and the dash 
			indicates approximate calculations (Eq.~(\ref{PQE})) for d=5,8,15~nm NCs and the bulk using the effective dephasing
			rates $\hbar\gamma_{eff}=50,40,55$, and 45 meV, respectively. (b) Solid lines from (a) plotted on the unitless
			pump energy scale (i.e., normalized by corresponding NCs and bulk band gap energies, $E_g$).} 
\label{qeth}
\end{figure}

The effective Coulomb as a function of energy is plotted in Fig.~\ref{CDOS}~(c) and follows power law behavior, $V_{eff}^{x,xx}(\omega)\sim\omega^{-5}$. According to the plot, the effective Coulomb interaction is only  
weakly enhanced by quantum confinement, except for small, $d=5$~nm, NCs. Specifically, the confinement results in a factor two enhancement in NCs with $d=8$~nm as compared to the bulk. Therefore, the scaling of the actual interband Coulomb interaction, i.e. $V_{C}=(v/V)^2V_{eff}^{x,xx}$, in NCs (inset to Fig.~\ref{CDOS}~(c)) in the region of high DOS should be dominated by the volume prefactor, i.e.,  $\sim d^{-6}$. However, we remind that the volume prefactor responsible for such dramatic scaling does not enter the quantities determining the QE except the Auger recombination rate (Eq.~(\ref{NxE})--(\ref{kAR})).   

\subsection{Photogeneration QE}
\label{pge}

The photogeneration QE as a function of the absolute photon energy in NC and bulk PbSe is shown in Fig.~\ref{qeth}(a). According to the plot, the QE at fixed photon energy monotonically increases with the NC diameter but does not exceed the bulk value. In contrast, the QE plotted on the unitless photon energy scale (Fig.~\ref{qeth}(b)) shows the opposite trend useful for photovoltaic applications. However, the absolute energy scale representation is more suitable to understand the physical mechanisms of photogeneration and, therefor, we continue our analysis using this scale. First, we compare our calculations for small, $d=5$~nm, NC with the calculations reported by Silvestri and Agranovich\cite{silvestri10} for $d=6$~nm PbSe NC. In Ref.~[\onlinecite{silvestri10}], the size dispersion of the NCs is set to 2\% resulting in the sharp peaks reaching $QE\approx 1.2$. The increase in the dispersion to realistic value of $5\%$ (red solid line in Fig.~\ref{qeth}(a)) washes out the peaks and reduces the QE by one order of magnitude, i.e. to $QE\sim 1.01$.

\begin{figure}[t]
\begin{center}
\epsfig{file=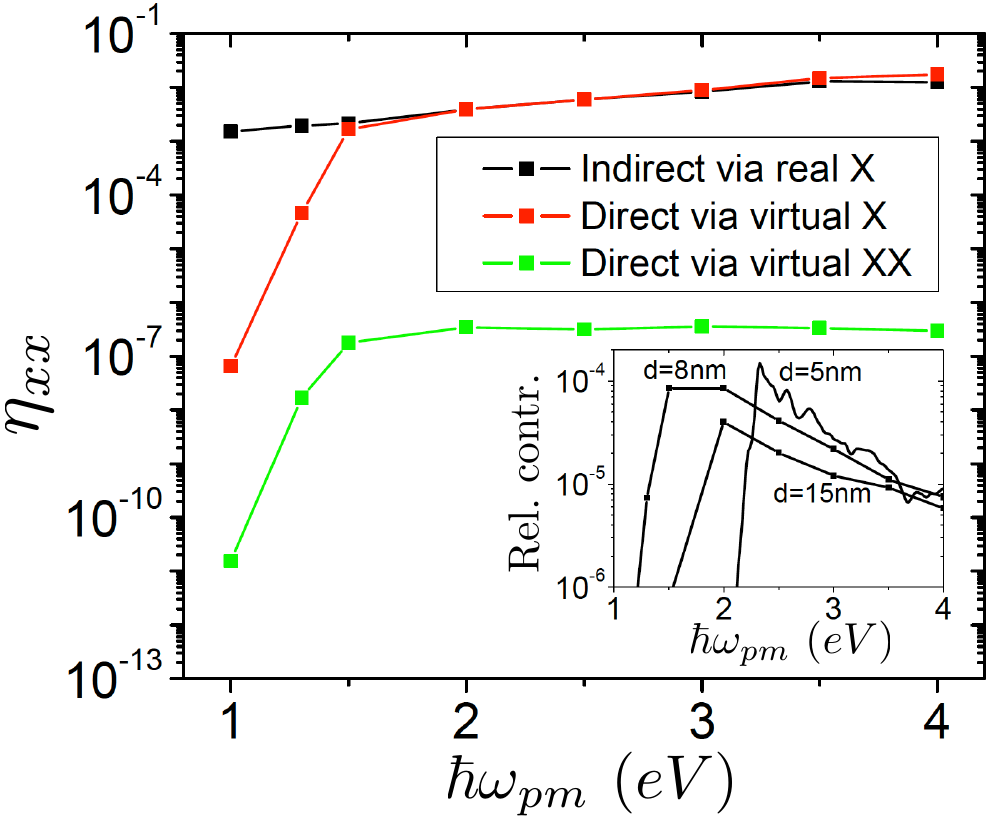,width=3.0in}
\end{center}
\caption{Contributions of the photogeneration pathways to the biexciton quantum yield calculated
		for $d=15$~nm PbSe NC. Black and red lines show the contributions of
		the two components associated with the first two pathways schematically shown in
		panels~(a) and (b) of Fig.~\ref{Fig-spath}, respectively. The green line shows
		the contribution of the third pathway illustrated in panel~(c) of Fig.~\ref{Fig-spath}
		The inset presents the relative contribution of the third pathway to the total photogeneration
		quantum yield in various size NCs.
		}
\label{qe_m1}
\end{figure}

The contributions of different pathways (Fig.~\ref{Fig-spath}) to the biexciton photogeneration quantum yield, $\eta_{xx}=QE-1$, are depicted in Fig.~\ref{qe_m1}. First of all, we point out that for the energy region $\hbar\omega_{pm}>1.5$~eV in which the biexciton DOS shows steep growth, the indirect biexciton photogeneration (black line) and direct biexciton photogeneration via exciton state  (red line), provide identical contributions. This behavior can be interpreted by looking at the related first and second terms in Eq.~(\ref{NxxE}) in which the resonant nature of the denominator, $(\omega^{'}-\omega_{pm})^2+\gamma^2$, corresponds to the leading contribution of the diagonal component of the effective Coulomb term $V^{x,xx}_{eff}(\omega_{pm})\equiv V^{x,xx}_{eff}(\omega_{pm},\omega_{pm})$. Hence, the integral convolutions can be evaluated resulting in the identical two terms whose net contribution to the biexciton population is
\begin{eqnarray}\label{NxxE1}
\tilde N^{'}_{xx}(\omega_{pm}) = k_{II}(\omega_{pm})\tilde N_{x}(\omega_{pm})/\gamma_{eff}.
\end{eqnarray}
Here, $k_{II}$ is the II rate given by Eq.~(\ref{kII}), and $\gamma_{eff}$ is effective interband dephasing rate. Derived expression has a very clear physical interpretation: {\em The photogenerated biexciton population associated with the first and second pathways is a result of a single II event taking place on the dephasing timescale, $\gamma_{eff}^{-1}$, and following optical preparation of the exciton states.}\footnote{For $\tau_{pm}>\gamma_{eff}^{-1}$, the $\gamma_{eff}^{-1}$ characterizes the timescale of the pump pulse interaction with NCs.}   

The direct biexciton photogeneration via exciton states is manifested by the direct dependence of the biexciton DOS on $\omega_{pm}$ in the second term of Eq.~(\ref{NxxE}). As a result, for $\hbar\omega_{pm}<1.5$~eV, this contribution (red line in Fig.~(\ref{qe_m1})) drops quickly. Accordingly, $\hbar\omega_{pm}=1.5$~eV can be identified as the AET for the photogeneration processes in $d=15$~nm NCs. In contrast, the indirect biexciton photogeneration (black line in Fig.~(\ref{qe_m1})) is resonant at exciton states, and decreases relatively slow for  $\hbar\omega_{pm} \lesssim 1.5$~eV. According to the first term in Eq.~(\ref{NxxE}), this behavior reflects the behavior of the off-diagonal effective Coulomb component $V^{x,xx}_{eff}(\omega^{'},\omega_{pm})$ at $\hbar\omega^{'}\lesssim 1.5$~eV.

\begin{figure}[t]
\begin{center}
\epsfig{file=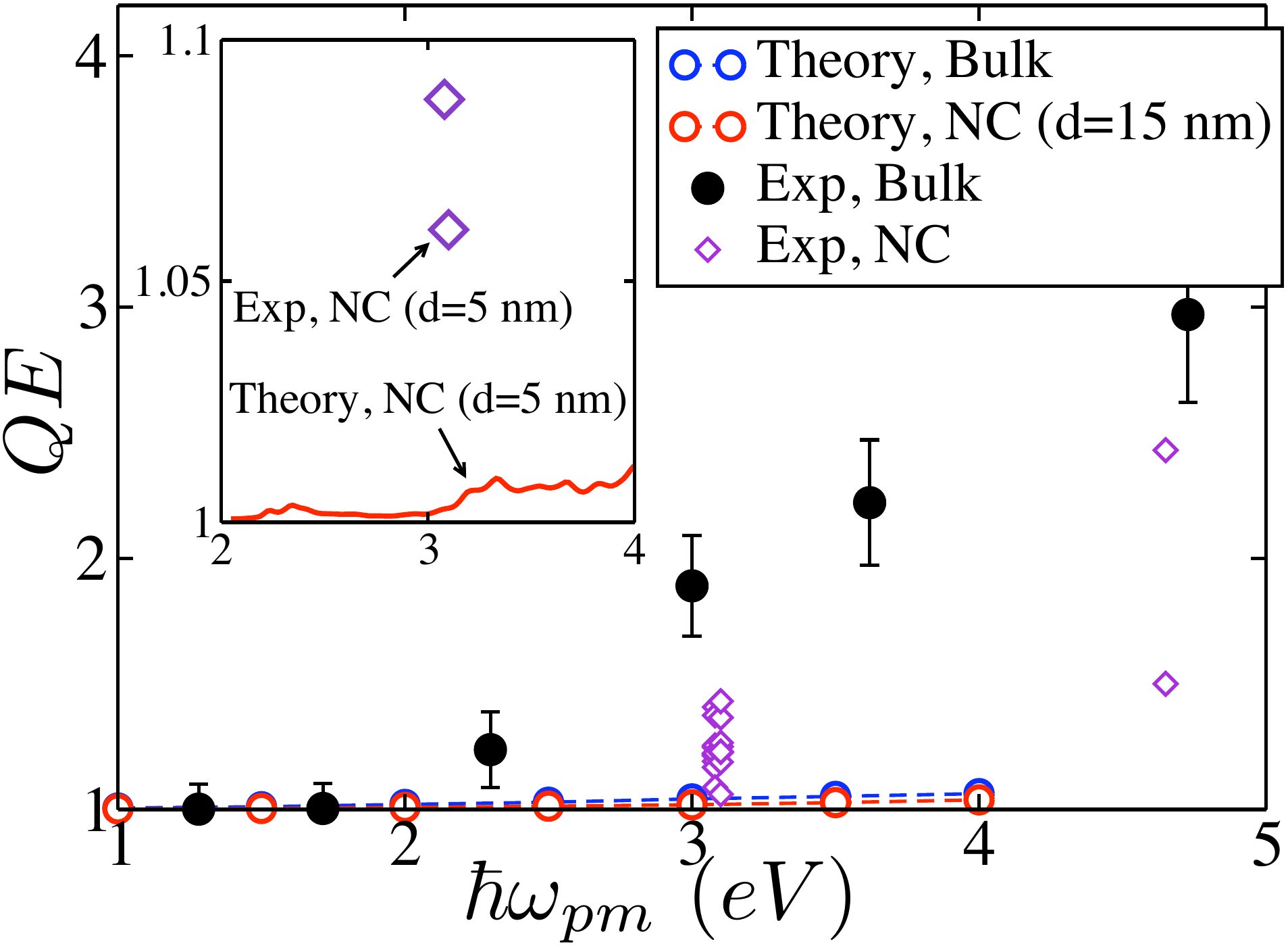,width=3.05in}
\end{center}
\caption{Calculated photogeneration QE as a function of pump energy for PbSe NC of $d=15$~nm and PbSe bulk. 
			For comparison, we also show experimentally measured total QE in PbSe NCs\cite{mcguire08} with the 
			diameter varied in the range of $5\leq d\leq 8$~nm and in the bulk\cite{pijpers09}. The inset shows 
			calculated QE for PbSe NC of $d=5$~nm on extended pump energy scale compared to the experimentally 
			measured total QE for the same size NC.}
\label{qe_exp_theory}
\end{figure}

According to Fig.~\ref{qe_m1}, the direct biexciton photogeneration via biexciton states (green line) has negligibly small contribution compared to the other terms. The inset shows that the relative contribution (i.e., normalized per total photogeneration QE) of the latter pathway for all considered NC sizes is small and has tendency to decrease with the NC size growth and to vanish in the bulk limit.  This contribution becomes small due to the strong off-resonant nature of the ground state to biexciton state effective Coulomb term (Eq.~(\ref{Veff0})) entering the third term in Eq.~(\ref{NxxE}).  

The small contribution of the last pathway leads us to an important conclusion that {\em the photogeneration QE is fully determined by a single II event occurring on the dephasing timescale.} According to Eqs.~(\ref{QE}) and (\ref{NxxE1}) this QE can be approximated by the following simple expression 
\begin{eqnarray}\label{PQE}
QE(\omega_{pm}) = \frac{2k_{II}(\omega_{pm})+\gamma_{eff}}{k_{II}(\omega_{pm})+\gamma_{eff}}\approx 1+\frac{k_{II}}{\gamma_{eff}},
\end{eqnarray}
where $k_{II}$ is the II rate given by Eq.~(\ref{kII}), and $\gamma_{eff}$ is effective interband dephasing rate. Note that the optical selection rules do not enter Eq.~(\ref{PQE}). The approximate form of QE given in Eq.~(\ref{PQE}) is quite general, since the approximation is based on a general fact of weak interband Coulomb interaction. To verify this relation, we calculated the QE using Eqs.~(\ref{kII}) and (\ref{PQE}), and compare the results (dashed lines) in Fig.~\ref{qeth}(a). In general, the dashed lines well reproduce the trends in the behavior of the associated solid lines confirming the validity of the approximation. Observed small discrepancies are due to the numerical noise, phenomenological origin of the effective dephasing rate neglecting its frequency dependence, and neglect of the correlations between the effective Coulomb and DOS fluctuations during the NC ensemble averaging.

Both {\em direct} biexciton photogeneration pathways via exciton and biexciton states have been studied before.\cite{schaller05,rupasov07,silvestri10} The {\em indirect} biexciton generation as we demonstrated above is also important and should not be omitted. Initial studies of the photogeneration pathways\cite{schaller05,rupasov07} also based on the KW parameterization reported significantly larger QE compared to our results and the results report in Ref.~[\onlinecite{silvestri10}]. The cause of the overestimate is the disregard of the Coulomb coupling selection rules and the oscillator strengths factors weighting optically allowed transitions.\cite{silvestri10} Next, we show that the interference related power-scaling of the DOS not accounted for in the previous studies further leads to significant reduction in the photogenerated QE.

Provided, the terms in Eq.~(\ref{nxx}) associated with the direct biexciton photogeneration processes via exciton and biexciton states are interfering {\em constructively}, one can show that associated biexciton populations scale as $[\rho^{'}_{x}]^2\rho_{xx}$ and $[\rho^{'}_{xx}]^2\rho_{xx}$, respectively. This is exactly the case considered in Refs.~[\onlinecite{schaller05,rupasov07}] Here, $\rho_{x}^{'}$ ($\rho_{xx}^{'}$) is DOS for the intermediate exciton (biexciton) states, and $\rho_{xx}$ the biexciton DOS at the pump frequency. Since, we found that there is {\em no interference} between the photogeneration pathways, the first two terms in Eq.~(\ref{NxxE}) scale as $\rho^{'}_{x}\rho_{xx}$, and the last one as $\tilde\rho_{xx}\sim{\rho_{xx}}^{'}\rho_{xx}$. The latter linear dependence on the intermediate exciton and biexciton DOS versus former quadratic one\cite{schaller05,rupasov07} significantly decreases the biexciton photogeneration quantum yield and, in fact,  allows us to interpret the photogeneration dynamics as a single II event (Eq.~(\ref{NxxE1}))!

In Fig.~\ref{qe_exp_theory}, we compare the experimentally measured total QE\cite{mcguire08,pijpers09} with the calculated photogeneration QE. The comparison with experimental data for NCs is possible only at the photon energy fixed at $\sim 3.1$~eV, where the vertical aligned  diamonds represent the QE measured for the NCs with diameters varying in the range between  $5$~nm and $8$~nm. The comparison shows that for the bulk and for the NCs of comparable diameter the experimentally measured total QE significantly exceeds the calculated photogeneration QE. Accordingly, we conclude that the biexciton photoexcitation along does not explain the experimentally  observed QE values. Next, we investigate the contribution of II events to the total QE taking place during the phonon-assisted cooling.

\begin{figure}[t]
\begin{center}
\epsfig{file=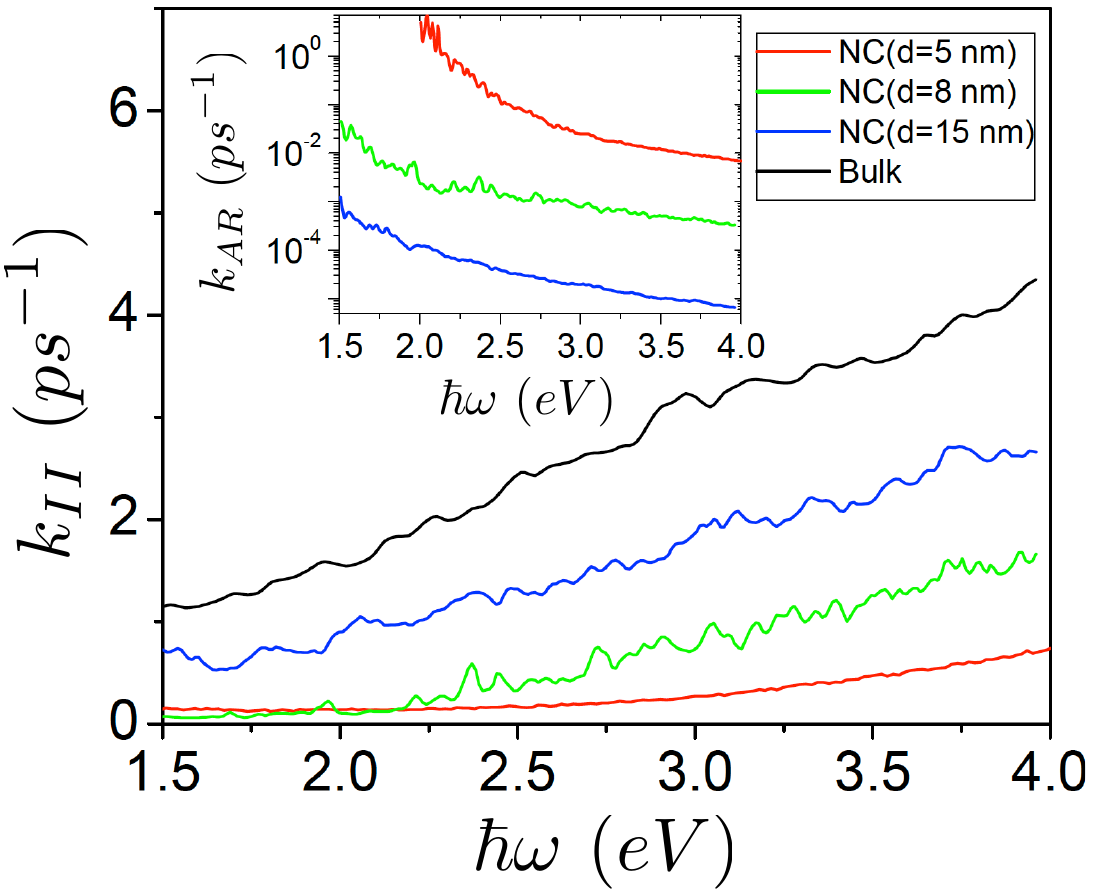,width=3.0in}
\end{center}
\caption{Energy dependence of the calculated II rates in PbSe NCs of different diameter and in PbSe bulk. 
			The inset shows associated Auger recombination rates calculated for NCs using (Eq.~(\ref{kr})).}
\label{iirates}
\end{figure}

\subsection{Effect of population relaxation and pump pulse duration on total QE}

Central quantity defining QE during the population relaxation (and as concluded above during the photogeneration event as well) is the II rate (Eq.~(\ref{kII})). Fig.~\ref{iirates} shows calculated energy dependence of the II rate for NC and bulk PbSe. According to the plot, the rate scales linearly away from AET. The scaling directly follows from Eq.~(\ref{kII}) in which one can insert the effective Coulomb, $V_{eff}^{x,xx}(\omega)\sim\omega^{-5}$, and biexciton DOS, $\rho_{xx}(\omega)\sim (\omega-2\omega_g)^{6}$, whose scalings  were determined at the beginning of this section (Fig.~\ref{CDOS}). In the literature, quadratic and even higher power behaviors of the II rate are obtained {\em near} the AET. As proposed in Refs.~[\onlinecite{keldysh65,ridley87,landsberg03}] such scalings solely follow from the DOS behavior. However, these theories do not account for the long-range interband Coulomb corrections at the energies higher than AET. As we show in Appendix~\ref{Appx_mtxl}, the first non-vanishing ${\bf k\cdot p}$ contribution to the matrix elements of the interband Coulomb term yields additional factor $\omega^{-1/2}$. As a result, one gets $k_{II}(\omega)\sim (\omega-\omega_0)^2/\omega\sim\omega$ at $\omega\gg\omega_0$, which is in agreement with our numerical calculations.

\begin{figure}[t]
\begin{center}
\epsfig{file=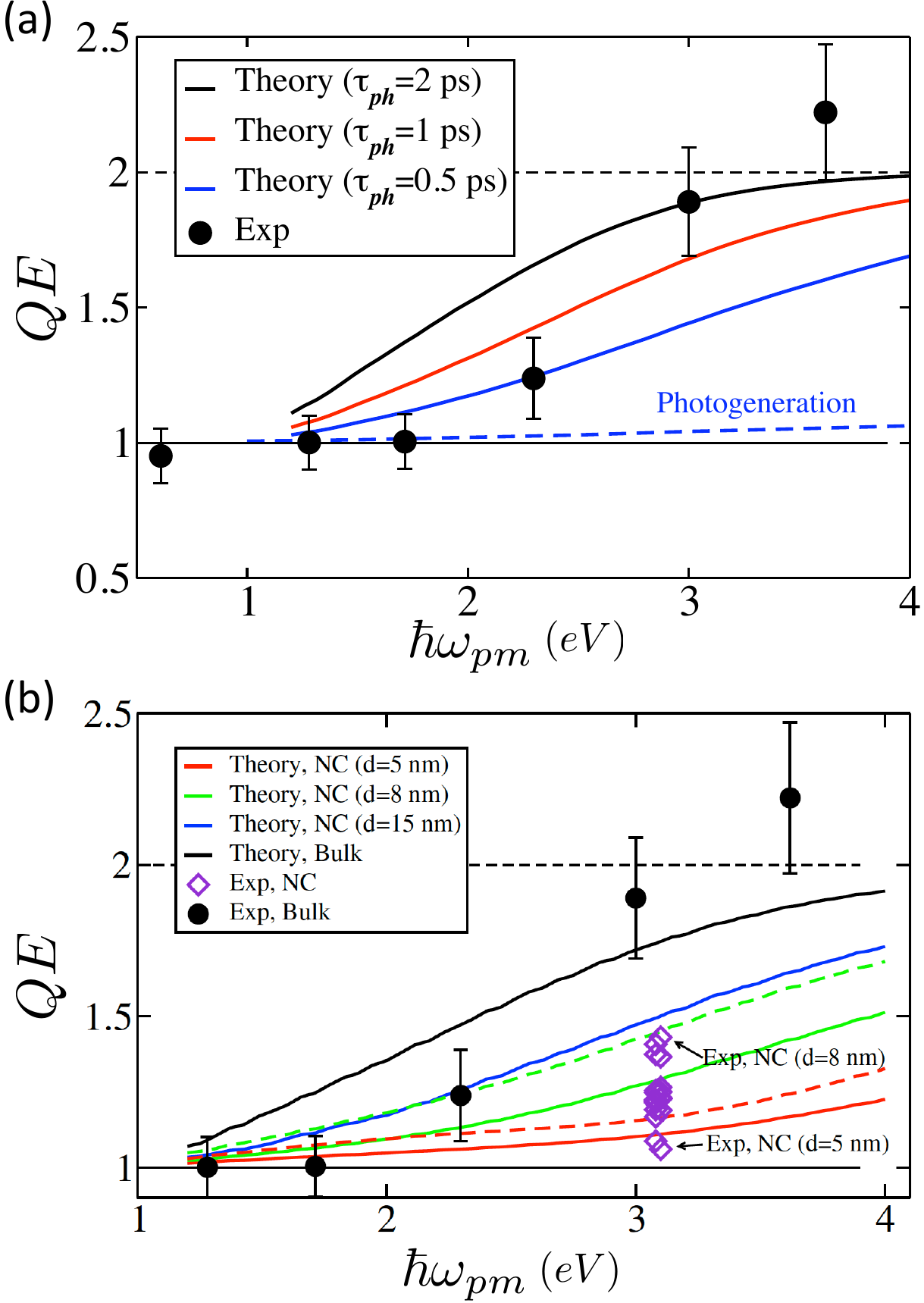,width=3.2in}
\end{center}
\caption{ Pump energy dependence of the total QE in PbSe. (a) Calculations performed for bulk PbSe using various intraband 
			relaxation times, $\tau_{ph}$. Experimental data from Ref.~[\onlinecite{pijpers09}] are shown for comparison.
			(b) Calculations for both NCs and bulk using $\tau_{ph}=1$~ps (solid lines) and $\tau_{ph}=2$~ps (dash). 
			For comparison the experimental data from Refs.~[\onlinecite{mcguire08}] and [\onlinecite{pijpers09}] are 
			plotted in diamonds and dots, respectively.}
\label{QE-total}
\end{figure}

According to Fig.~\ref{iirates}, the II rate increases as the NC diameter increases but does not exceed the bulk values. This observation is a result of the {\em weak} effective Coulomb enhancement in NCs observed in Fig.~\ref{CDOS}~(c). In agreement with previously reported studies,\cite{pijpers09} we find that this enhancement is fully suppressed by the reduction in the biexciton DOS (Fig.~\ref{CDOS}~(b)). As we already pointed out, strong size scaling of the actual interband Coulomb interaction is determined by the volume prefactor $(v/V)^2$ which does not enter the II rate, and therefore, has no effect on the rate.  

In contrast, the volume prefactor appears in the expression for the Auger recombination rate (Eq.~(\ref{kAR})), and as we show in the inset to Fig.~\ref{iirates}, makes this effect negligible. On the frequency scale, the Auger recombination rate drops as $k_{AR}\sim\omega^{-3}$ further decreasing its contribution to QE at high energies. As a result, the only region where the Auger recombination processes are significant is near the band edge. However, this region has no significant contribution to QE, and we conclude that the Auger recombination processes can be fully neglected in the present analysis. 

\begin{figure}[t]
\begin{center}
\epsfig{file=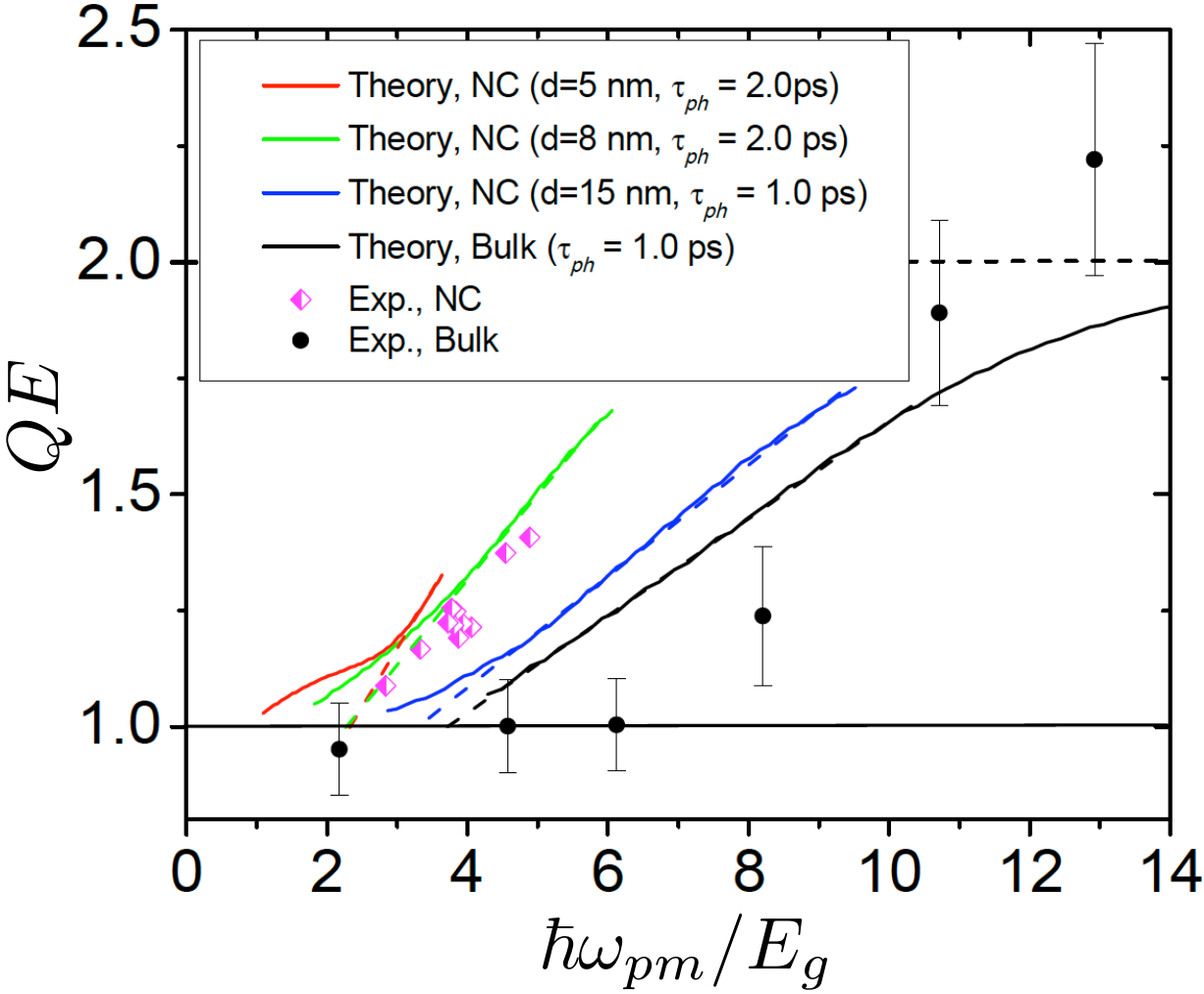,width=3.4in}
\end{center}
\caption{Comparison of the calculated total QE for selected $\tau_{ph}$ and experimental data from Fig.~\ref{QE-total}~(b)
plotted on the unitless pump energy scale (i.e., normalized per corresponding band gap energy, $E_g$). Dashed lines
are extrapolation of the solid lines linear regions to determine the AET values (see Table~\ref{Table-AET}). 
}
\label{QE-uless}
\end{figure}

Since the Auger recombination processes are weak, efficient CM during the population relaxation is expected if the II rates are comparable to or higher than the phonon-assisted population relaxation rates. To evaluate the upper boundary for QE in NCs, we first calculate total QE in bulk PbSe for typical population relaxation times $\tau_{ph}=$~0.5, 1.0, and 2~ps. The results are plotted on the pump energy scale in  Fig.~\ref{QE-total}~(a) and also compared with published experimental data. In general, the calculations reproduce the experimental trends and provide the best fit at $\tau_{ph}=1.0$~ps. The discrepancy between theory and experiment in the interval $1.0< \hbar\omega_{pm}<3.5$~eV is due to the phenomenological nature of the phonon-assisted relaxation model that lacks exact knowledge on the spectral dependence of the electron-phonon coupling and phonon DOS. Above $3.5$~eV the effect of triexciton generation and possibly the contribution from the higher energy bands take place. Hence, our theory is valid below this energy value.    

\begin{table}[b]
\begin{center}
\begin{tabular}{|c|c|c|c|c|}
 \hline
 PbSe & $d$~(nm) & $E_g$~(eV) & AET/$E_g$ &AET~(eV) \\\hline
 NC & 5 & 1.10 & 2.3 &2.5 \\
 NC & 8 & 0.66 & 2.2 &1.5 \\
 NC & 15 & 0.42 & 3.3 &1.4  \\
 Bulk & $\infty$ & 0.28 & 3.7 & 1.0 \\\hline
\end{tabular}
\end{center}
\caption{AET in PbSe NCs and bulk calculated as the intercept point between 
dashed lines and the solid horizontal line at QE=1 in Fig.~\ref{QE-uless}.}
\label{Table-AET}
\end{table}

For comparison, the contribution of the photogeneration to the total QE is shown in Fig.~\ref{QE-total}~(a).  Its small contribution is explained by the fact that the photogeneration pathways reduce to a {\em single} II event taking place on the short (subpicosecond) dephasing timescale (Eq.~(\ref{PQE})). In contrast, the population relaxation dynamics occurs on a much longer (picosecond) timescale allowing for the {\em multiple} II events that make major input into the total QE.

Comparison of the calculated total QE in both NC and bulk PbSe with experimental results\cite{mcguire08,pijpers09} is shown in Fig.~\ref{QE-total}~(b). The total QE dependence on the NCs diameter  both in theory and experiment follows the same trend as the II rate. Solid lines in Fig.~\ref{QE-total}~(b), show QE calculated for $\tau_{ph}=1.0$~ps which gives the best fit for the bulk. However, taking into account that the upper values of the measured QE in the NCs are associated with the NCs diameter of $d=8$~nm, we had to increase the phonon relaxation time up to $\tau_{ph}=2$~ps for the NC with $d=8$~nm (blue dash) in order to get better agreement with experiment. The observed increase of $\tau_{ph}$ can be rationalized by the quantum-confinement-induced increase of the level spacing.\cite{bonati07} 

Determination of AET is more convenient to perform using the unitless energy scale, i.e., the pump energy normalized per corresponding $E_g$ of NC or bulk. Fig.~\ref{QE-uless} presents the same curves as in Fig.~\ref{QE-total}~(b) for selected $\tau_{ph}$ plotted on this scale. Dashed lines extrapolate the linear portion of the curves at the energies below $3.5$~eV. Their intercept points with the horizontal, $QE=1$, line provide the AET values. The deviation from linear behavior at low energy ends and apparent CM below the energy conservation threshold, $2E_g$ (red and green curves) are merely due to the ensemble averaging entering our calculations.\footnote{Specifically, the ATE has some energy distribution associated with the $5\%$ diameter variations. However, the $E_g$ used to normalize the photon energy is an averaged value of the band gap. Therefore, the values of QE below {\em averaged} $2E_g$ are due to the contribution of the sub-ensemble with actual $E_g$ values below its mean value.}      

The calculated values of AET are summarized in Table~\ref{Table-AET}. For the NCs with $d=5$~nm and $d=8$~nm, the calculated AET$\approx 2.2E_g$ is in excellent agreement with experiment.\cite{mcguire08} Furthermore, the {\em photogeneration} AET=1.5~eV for $d=15$~nm PbSe NCs (determined in Sec.~\ref{pge}, Fig.~\ref{qe_m1}) is close to that, $1.4$~eV, listed in Table~\ref{Table-AET}. The latter is an additional indication of the ``pure" II nature of CM. For the bulk however, there is a disagreement between the calculated, $3.7E_g$, and experimentally observed, $6E_g$, AET.\cite{pijpers09} The discrepancy is most likely related to the phenomenological model of the phonon-assisted relaxation. Getting back to Fig.~\ref{QE-total}~(a), one can clearly see that the best fit to the initial rise of QE can be obtained for $\tau_{ph}=0.5$~ps.  However, for adopted $\tau_{ph}=1$~ps the model fails to accurately reproduce the initial slope resulting in the underestimated values of AET.    

\begin{figure}[t]
\begin{center}
\epsfig{file=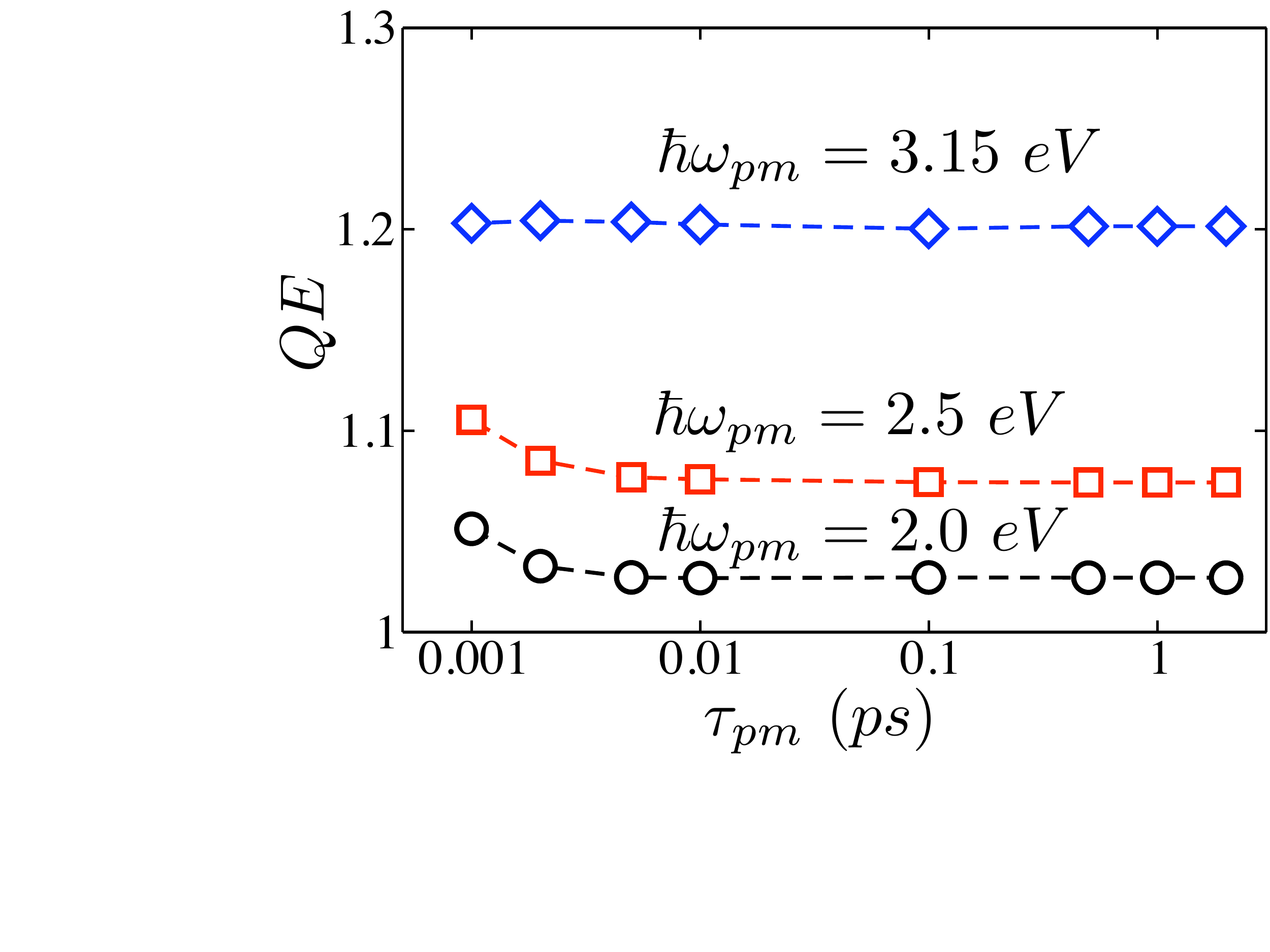,width=3.2in}
\end{center}
\caption{\label{diff_tauph} Calculated dependence of the total QE for PbSe NC of $d=8$~nm on the pump pulse duration $\tau_{pm}$. 
			The relaxation time is set to $\tau_{ph}=1$~ps.}
\end{figure}

All calculations discussed above are performed for the cw pulses typically used in ultrafast spectroscopic studies. However, estimated solar light correlation time is about $5$~fs. Hence in light of photovoltaic applications, it is natural to calculate QE as a function of the pump pulse duration, $\tau_{pm}$.  Such a dependence evaluated for various photon energies, $\hbar\omega_{pm}$, is shown in Fig.~\ref{diff_tauph}. First of all, we notice that the cw regime is reached at $\tau_{pm}>10$ fs. For $\tau_{pm}<10$~fs, the QE associated with $\hbar\omega_{pm}=3.15$~eV shows no variation whereas QE associated with the lower excitation energies show insignificantly small increase. Observed weak dependence of QE on the pulse duration suggests that, for the gaussian pulses, the expected increase in the QE due to the increase in the pulse duration is equally compensated by the reduction in the number of states populated by the pulse whose spectral widths is narrowed.

\section{ Comparison of QE in P\lc{b}S and P\lc{b}S\lc{e}}
\label{comp}

\begin{figure}[t]
\centering
\includegraphics[width=2.8in]{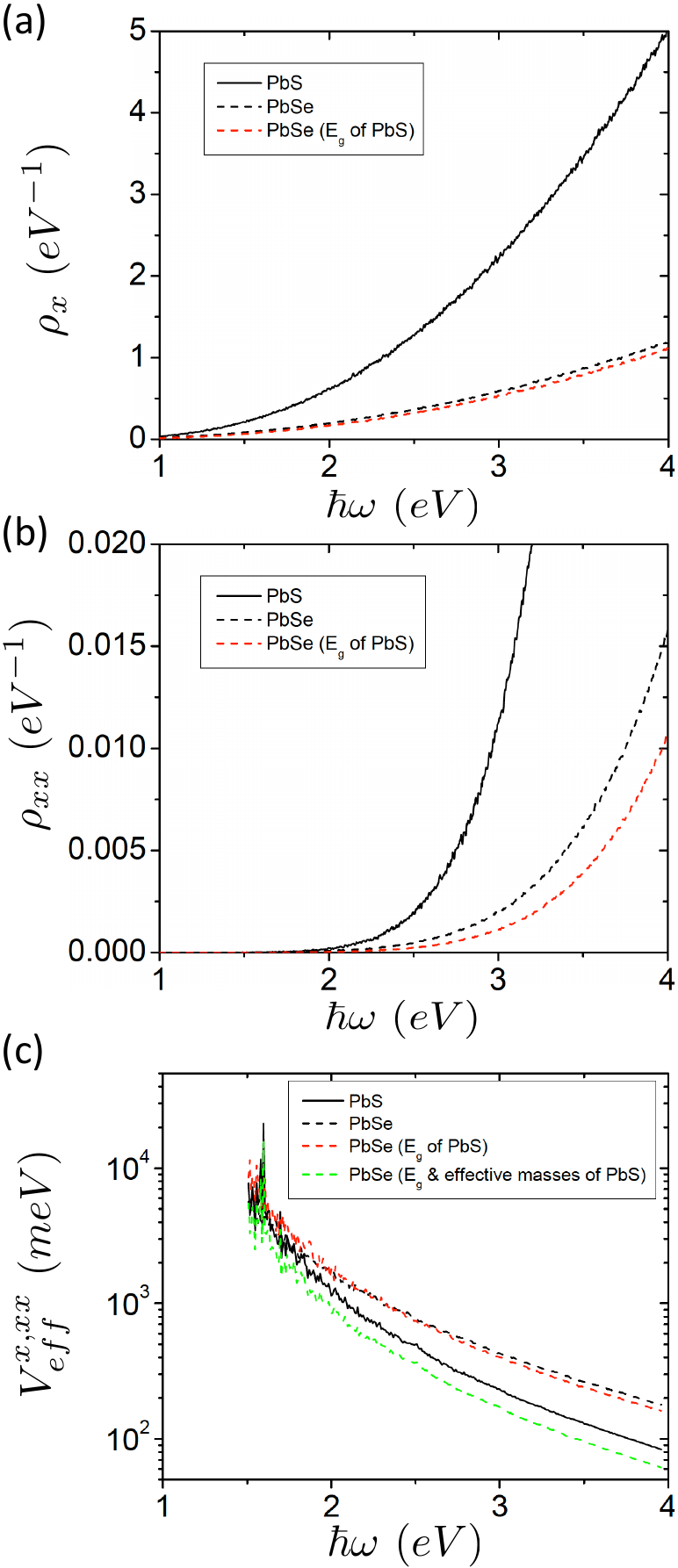}
\caption{Energy dependence of (a) exciton DOS, (b) biexciton DOS, and (c) effective Coulomb term in the bulk PbS 
	(solid black) and PbSe (dashed black). In all panels, red lines show corresponding quantities calculated using PbSe parameters 
	except for the band gap energy replaced with that of PbS. In panel~(c), green line shows the effective Coulomb term calculated 
	for PbSe with both band gap energy and the effective masses substituted with their corresponding PbS values.}
\label{dos_comparison}
\end{figure}

Both PbS and PbSe have electronic structures described by the same KW effective mass model with some different parameters.\cite{kang97} This implies that our conclusions on the interplay between different CM pathways in PbSe fully apply to PbS NCs and bulk materials. In this section, we perform comparison of the key quantities such as exciton/biexciton DOS, effective Coulomb terms, the II rates, and the resulting QEs calculated for these semiconductors. Our numerical tests show that these quantities are weakly affected by the differences in the Kane momentum and the unit cell volume. Hence, the dominant contributions come from the interplay between the band gap energies, carriers effective masses, and dielectric constants.   

Panels~(a) and (b) in Fig.~\ref{dos_comparison} clearly show that both the exciton (Eqs.~(\ref{XDOS})) and biexciton (Eq.~(\ref{XXDOS})) DOS in bulk PbS exceed the corresponding DOS in PbSe across the whole spectral range of interest. Compared to PbS ($E_g=0.28$~eV), PbSe has larger band gap energy ($E_g=0.41$~eV) and heavier carriers effective masses.\cite{kang97} In order to isolate the band gap effect, we plot the PbSe DOS calculated with the band gap energy replaced by that of PbS (red dash). This replacement results in  a reduction of the DOS values (red dash). Hence, the major contribution to the steeper growth of the exciton and biexciton DOS in PbS comes from the heavier effective masses.    

\begin{figure}[t]
\centering
\includegraphics[width=3.2in]{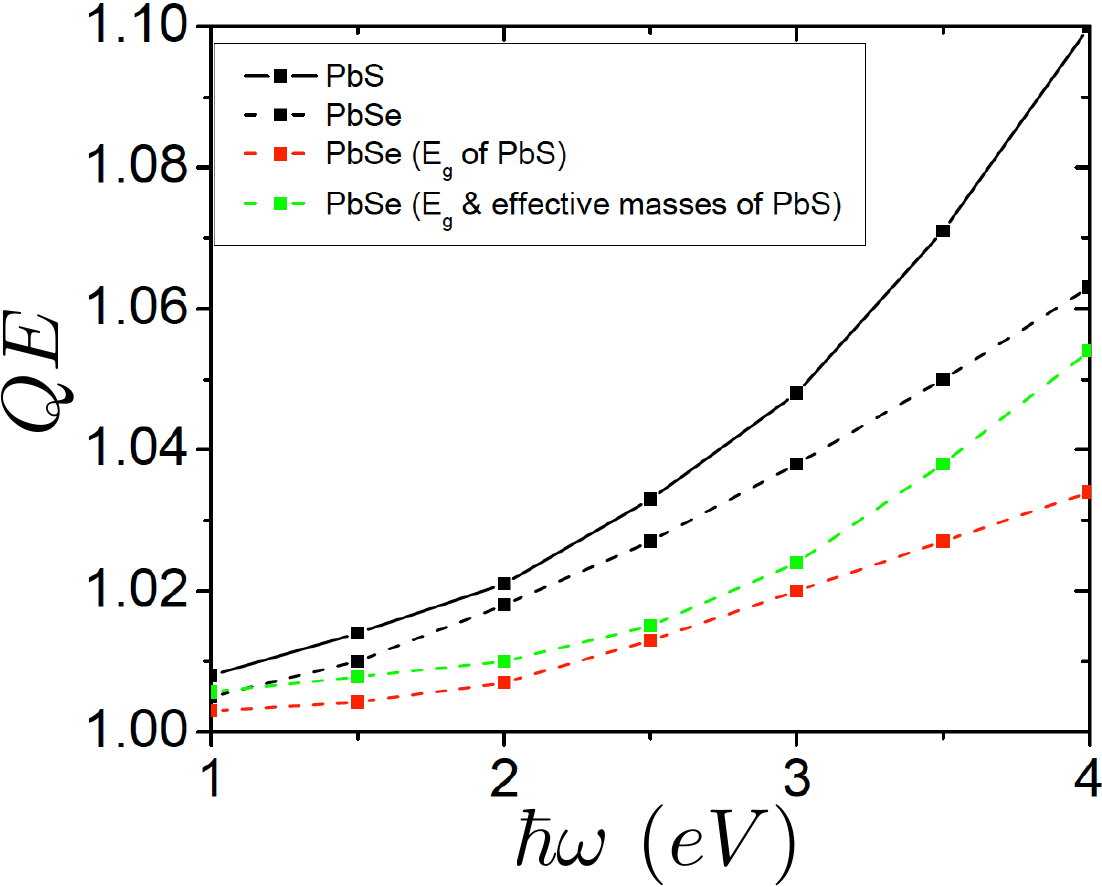}
\caption{Energy dependence of the photogeneration QE in the bulk PbS (solid black) and PbSe (dashed black). Red (green) line shows
		the QE calculated for PbSe with the band gap energy (band gap energy and the effective masses) replaced with that (those) of 
		PbS.}
\label{PQE-comp}
\end{figure}

Fig.~\ref{dos_comparison}~(c) compares the effective Coulomb terms, $V^{x,xx}_{eff}$, (Eq.~(\ref{Veff})) calculated for bulk PbSe and PbS. In contrast to the DOS, the effective Coulomb interaction in PbS is weaker than that in PbSe. To understand the trends, we alter the PbSe band gap value in the same way as for the DOS, and observe small reduction in $V^{x,xx}_{eff}$ (red dash). Altering the effective masses of PbSe with those of PbS, significantly reduces the $V^{x,xx}_{eff}$ lowering the curve (green dash) below $V^{x,xx}_{eff}$ in PbS (solid black). The gap between these curves is purely due to the difference in the dielectric constants which in PbS ($\epsilon=17$) is lower  than in PbSe ($\epsilon=23$). Therefore according to Fig.~\ref{dos_comparison}~(c), the narrower band gap value and lighter effective masses make dominant contribution $V_{eff}^{x,xx}$ in PbS making it smaller than in PbSe.     

\begin{figure}[t]
\centering
\includegraphics[width=3.2in]{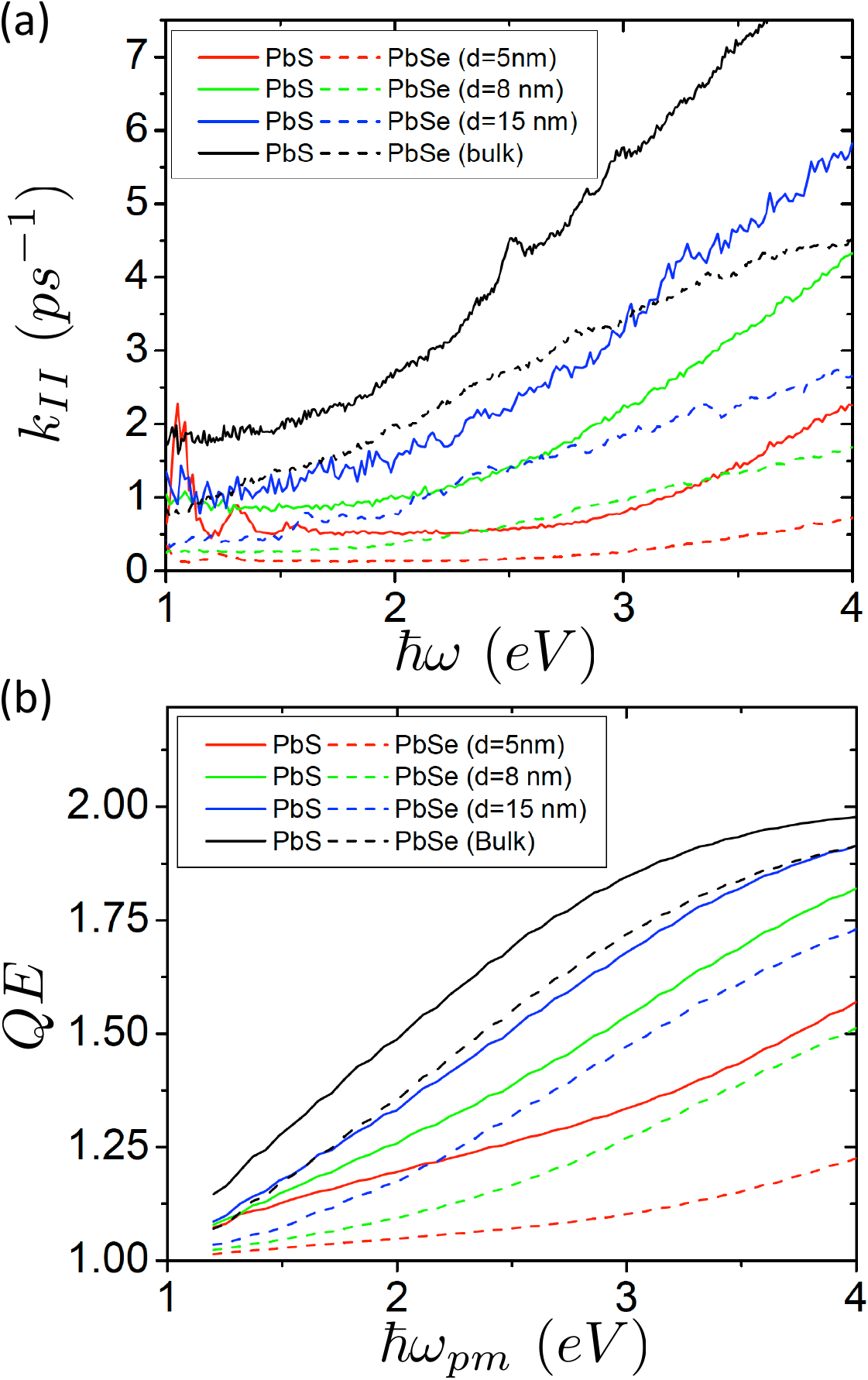}
\caption{Energy dependence of (a) II rate and (b) total QE calculated for PbS (solid lines) and PbSe (dash) NCs, 
		and the bulk. For all curves in panel ~(b), the phonon relaxation time is set to $\tau_{ph}=1.0$~ps.}
\label{kII-comp}
\end{figure}

Next, we compare the photogeneration QE  in bulk PbS and PbSe plotted in Fig.~\ref{PQE-comp}. The plot also shows photogeneration QE values for PbSe with the band gap energy and effective masses altered in the same way as in the case of the  effective Coulomb interaction. (The color code is the same as in Fig.~\ref{dos_comparison}~(c).) By taking into account the linear dependence of the photogeneration QE on the II rate (Eq.~(\ref{PQE})), one can conclude that the trends in the behavior of the QE is a result of the interplay between the trends in the biexciton DOS and effective Coulomb term shown in panels~(b) and (c) of  Fig.~\ref{dos_comparison}, respectively. Specifically, the lower dielectric constant of PbS contributes to the increase  in the photogeneration QE and the competing effect of the lighter effective masses on the DOS and $V_{eff}^{x,xx}$ results in additional increase of the QE in bulk PbS.  This net increase fully overruns the reductions associated with the band gap decrease making the photogeneration QE and the II rate (Fig.~\ref{kII-comp}~(a)) in bulk PbS to be higher than in bulk PbSe.     

To demonstrate that the size quantization does not change the trends, we plot the II rate calculated for PbS and PbSe NCs and the bulk in Fig.~\ref{kII-comp}~(a). It is clear from the plot that  the PbS II rate always exceeds the PbSe II for identical in diameter NCs and for the bulk. Similar to PbSe, the PbS II rate monotonically   increases with the NC diameter but does not exceed the bulk values. Finally in panel~(b)  of Fig.~\ref{kII-comp}, we plot the calculated total QE of PbS and PbSe NCs and the bulk using {\em the same} phonon relaxation time set to $\tau_{ph}=1.0$~ps. The trends are the same as in panel~(a) demonstrating that the overall {\em calculated} QE in PbS is higher than in PbSe. The latter is a result of the interplay between the band gap energies, effective masses and the dielectric constant values as discussed above.  

Ultrafast measurements of QE in bulk PbS and PbSe, show that QE for both materials are approximately the same.\cite{pijpers09} However, the atomistic calculations reported along with the experimental data demonstrate that for identical phonon relaxation times (specifically, $\tau_{ph}=0.5$~ps) the calculated QE in PbS exceeds that in PbSe. This observation is in direct agreement with our calculations. Taking into account that both models show the same trend in QE for identical phonon relaxation times, we conclude that the II rate in bulk PbS should be higher than in PbSe, and consequently, the photogeneration QE. Furthermore, we extend this conclusion to PbS NCs according to the results shown in Fig.~\ref{kII-comp}. 

Recently reported ultrafast measurements demonstrated significant reduction in QE of large PbS NCs compared to PbSe NCs.\cite{stewart12} The authors rationalize this observation by estimating the phonon-induced energy loss rate which in PbS turns out to be factor two faster compared to PbSe. Rigorous calculation of the phonon-induced relaxation is beyond our model capability forcing us to introduce phenomenological $\tau_{ph}$. Since remaining electronic structure parameters in our and reported atomistic models are well validated, the calculations directly support the idea of the QE reduction in PbS merely due to the fast phonon-induced relaxation processes characterized by $\tau_{ph}$. 

\section{Discussion and Conclusions}
\label{disc}

The effective mass model adopted in this paper has significant limitations accounting only for the $L$-valley optical transitions.\cite{kang97} It fails to catch contributions from the higher energy transitions originating at other Brillouin zone points. Specifically in PbSe, the $\Sigma$-point transitions show large contribution to the absorption spectra above $1.6$~eV.\cite{koole08} Experimentally, the CM dynamics is probed using ultrafast transient absorption and fluorescence techniques.\cite{mcguire08} These  techniques explicitly determine the number of carriers accumulated near the band edge of the lowest in energy $L$-valley. Following the excitation with a high energy optical pump, the excitons and biexcitons can be generated not only in $L$-valley (denoted $L$-excitons and biexcitons) but also at $\Sigma$-point (denoted $\Sigma$-excitons and biexcitons). The contribution of the $\Sigma$-excitons to the transient ultrafast signal depends on the mixing between the $L$ and $\Sigma$-points. Rigorous calculations of the mixing can be done through atomistic calculations. However, such calculations become tremendously expensive for the considered large diameter NCs. Below, we provide quantitative discussion of possible mixing mechanisms and rough estimates of the their contribution to QE for the pump energy below $3.5$~eV.

First, we consider mixing between valleys due to the size quantization which could be efficient in very small ($d\lesssim 3$~nm) NCs at low excitation energies. However, for the large NCs, considered here, the excitations with the energies higher than $2.5$~eV are effectively of the bulk-type\cite{koole08,moreels08} resulting in the negligible quantum-confinement-induced mixing. Another option is the II event initiated by the optically prepared $\Sigma$-exciton and resulting in the direct production of an extra $L$-exciton. Specifically, the excess energy of the $\Sigma$-exciton is transferred to create an extra electron-hole pair within $L$-valley. The AET for this process can be roughly estimated as $AET_{\Sigma L}\approx AET_L+(E_g^\Sigma-E_g^L)=3.3$~eV pointing to an {\em efficient} CM at photon energies $\hbar\omega_{pm}\gtrsim 4$~eV. Note that efficient CM events fully constrained to $\Sigma$-point should occur at even higher energies which are above the photon energy range considered in this paper. 

Finally, the optically prepared $\Sigma$-exciton can participate in the II event resulting in the creation of the electron-hole pair through the inter-valley transition. For instance, a higher energy conduction band electron from $\Sigma$-point can be transferred to $L$-valley  releasing in the excess energy to create an additional electron-hole pair through the $L$-valence band to $\Sigma$-conduction band transition.   As we mentioned above, most of the states involved in efficient CM are of the bulk-type with the quasi-momentum being a ``good" quantum number. Therefore, the Coulomb matrix elements describing the II processes become $\sim 1/|\Delta\bm q|^2$ where $\Delta\bm q$ is the difference between the electron/hole initial and final states. Accordingly, the II events involving the inter-valley transitions require large variation in the quasi-momentum, $\Delta\bm q$, and, therefore, becomes less favorable.

Experimental verification of the discussed interplay between the CM pathways contributing to the photogeneration QE requires direct measurements of the interband Coulomb interactions. Distinguishing the inter- and intraband Coulomb interactions is a challenging task, since in the transient absorption and time-resolved fluorescence experiments these two components contribute to a measured total energy shift between the exciton and biexciton bands. Recently, we have theoretically shown that the two-dimensional double-quantum coherence spectroscopy is capable to probe directly the interband Coulomb interactions.\cite{velizhanin11} Alternatively, the photogeneration QE can be determined based on its linear relationship to the II rate which has been established above (Eq.~(\ref{PQE})). The later rate can be obtained by fitting the total QE as a function of the excitation energy.\cite{beard10} However, the determination of the effective dephasing rate, also entering Eq.~(\ref{PQE}), might require use of coherent experimental technique (e.g., photon echo) and/or additional theoretical calculations. 

The absolute photon energy scale, $\hbar\omega_{pm}$, is used throughout this paper (except Fig.~\ref{qeth}(a) and \ref{QE-uless}). This scale is more useful to discuss fundamental physical mechanisms behind the CM dynamics, e.g., the role of the quantum confinement.\cite{nair08,delerue10} However, performance of photovoltaic devices can be characterized by various power conversion efficiencies.\cite{mcguire08,delerue10} In particular, by the QE calculated on a dimensionless energy scale which is  normalized per NC or bulk band gap, $\hbar\omega_{pm}/E_g$.  Despite lower values of QE in NCs compared to the bulk which show up on the absolute  photon energy scale, the performance of prospective photovoltaic devices  based on NCs can overrun their bulk counterpart.\cite{mcguire10,delerue10,beard10} By comparing Fig.~\ref{QE-uless} with Fig.~\ref{QE-total}~(b), one can notice that smaller size NCs have better rise in QE than large ones and the bulk. This results from the quantum-confinement-induced blue-shift of the band gap energy.\cite{nair11} In light of this effect and the requirement to match solar radiation peak energy, semimetal NCs can become more efficient solar energy converters.\cite{allan11}

Further enhancement of the interband Coulomb interactions in NCs can possibly be reached in NC-metal heterostructures in which broad surface plasmon response is tuned in resonance with the exciton states participating in CM. This issue requires additional study, since besides the enhancement of the Coulomb interactions, the surface-plasmons-induced ohmic and radiative energy losses can become significant.\cite{novotny2008} On the other hand, additional enhancement of QE can in principle be achieved by adding the surface states and ligands. This question is difficult to analyze using adopted effective mass approximation. However, the atomistic calculations\cite{kilina09a,allan09,hyeon11,albert11,fisher12,kilina12} combined with the IESM should be extremely helpful in addressing this problem. 

In conclusion, we have performed systematic numerical investigation of CM mechanisms in NC and bulk PbSe and PbS using our IESM parametrized by the effective mass KW model. We focused on the role of quantum confinement on the photogeneration and total QE. The analysis of the photogeneration pathways resulted in unexpected conclusion that the photogeneration processes reduce to a single II event due to the complete destruction of the associated pathways interference. This allowed us to explains the minor role of the photogeneration in total QE dominated by the multiple II events occurring during the phonon-induced population cooling. Comparison of the size scaling of the effective Coulomb interaction and the biexciton DOS in transition from NC to the bulk limit showed that weak enhancement of the former quantity is overrun by a significant reduction in the latter one. This explains the higher values of the QE in bulk compared to NCs as plotted on the absolute energy scale, and well agrees with the previously reported experimental studies and some theoretical predictions. However, the quantum confinement induced increase in $E_g$ makes NCs more efficient than bulk for practical photovoltaic applications. We have also found that the variation in the pump pulse duration does not significantly change the QE. Comparison of QE in PbSe and PbS suggests that the II processes are more efficient in PbS. However, variation of the material-dependent cooling time can strongly affect the total QE. Finally, we have identified the limitations of our model and defined its applicability range.  
 
\acknowledgements

K.A.V. acknowledge support of the Center for Advanced Solar Photophysics (CASP), an Energy Frontier Research Center funded by
the U.S. Department of Energy (DOE), Office of Science, Office of Basic Energy Sciences (BES). A.P. is supported by Los Alamos LDRD program. Both authors acknowledge CNLS and CINT for providing computational facilities, and wish to thank Victor Klimov, Darryl Smith, and Sergei Tretiak for stimulating discussions and comments on the manuscript.

\appendix

\section{Volume scalings of DOS, transition dipoles, and the effective Coulomb interaction}
\label{Appx_vscl}

In this Appendix, we derive the volume normalization prefactors entering Eqs.~(\ref{XDOS})--(\ref{Veff0}). The prefactors  cancel out the volume dependence of the latter quantities making them intensive variables in the bulk (i.e., thermodynamic) limit. To preserve the dimensionality of the intensive variables, we use the $V/v$ ratio instead of $V$ where $v$ is the unit cell volume. This ratio defines number of unit cells and goes to infinity in the bulk limit. Here, we also use the relationships derived in Appendix~\ref{Appx_mtxl} that connect the transition dipole movements and Coulomb matrix elements represented in the single-particle (KW) carrier basis set and in the exciton/biexciton basis.

We start with the simple fact that in the bulk limit, the single-particle DOS is proportional to the system volume V.\cite{kubo95I} Defining the exciton (biexciton) DOS as a joint DOS of an electron and a hole (two electrons and two holes), one immediately finds that they have $V^2$ ($V^4$) scalings. Therefore, we introduced $v^2/V^2$ ($v^4/V^4$) prefactor into Eq.~(\ref{XDOS}) (Eq.~(\ref{XXDOS})).

To obtain the volume prefactors in the optically allowed DOS given by Eqs.~(\ref{XODOS}) and (\ref{XXODOS}), we first notice that  the transition dipole matrix element for a single carrier has no volume scaling and accounts for the total momentum conservation, i.e., $M_{ij}\sim V^0 M_{{\bf k}_i,{\bf k}_j}\delta_{{\bf k}_i,{\bf k}_j}$.\cite{haug09} Further using Eq.~(\ref{mu_x}), we find that the optically allowed exciton DOS scales as
\begin{eqnarray}\label{VXODOS}
	&~&\sum_a|\mu^x_{a0}|^2\delta(\omega-\omega^x_a)\sim\\\nonumber&~&
		\sum_{{\bf k}_p}|M^{eh}_{{\bf k}_p,{\bf k}_p}|^2\delta(\omega-\omega^{e}_{{\bf k}_p}-\omega^{h}_{{\bf k}_p})\sim V.
\end{eqnarray}
Here and below, we use the same argument as in the DOS analysis that for a fixed energy interval, $\sum_{{\bf k}_p}\sim V$. Using the first term in the expansion of the intraband biexciton dipole moment given by Eq.~(\ref{mu_xx}), one finds that the joint optically allowed biexciton DOS scales as
\begin{eqnarray}\label{VXXODOS}
&~&\sum_{kl}|\mu^{xx}_{kl}|^2\delta(\omega_1-\omega^{xx}_k)\delta(\omega_2-\omega^{xx}_l)\sim
\\\nonumber&~&
\sum_{ {\bf k}_p{\bf k}_q}\sum_{ {\bf k}_r{\bf k}_s}|M^{ee}_{{\bf k}_p,{\bf k}_p}|^2
\delta(\omega_1-\omega^{e}_{{\bf k}_p} -\omega^{e}_{{\bf k}_q}-\omega^{h}_{{\bf k}_r}-\omega^{h}_{{\bf k}_s})
\\\nonumber&~&\times
\delta(\omega_2-\omega^{e'}_{{\bf k}_p} -\omega^{e}_{{\bf k}_q}-\omega^{h}_{{\bf k}_r}-\omega^{h}_{{\bf k}_s})
\sim V^4.
\end{eqnarray}
According to Eqs.~(\ref{VXODOS}) and (\ref{VXXODOS}), we introduced the prefactors $v/V$ and $(v/V)^4$ into Eqs.~(\ref{XODOS}) and (\ref{XXODOS}), respectively.

To determine the volume scaling of the effective Coulomb term (Eq.~(\ref{Veff})), we first evaluate the scaling of the following auxiliary quantity
\begin{equation}\label{VKaux}
\mathcal{K}(\omega_1,\omega_2)= \sum_{a,m}|V^{x,xx}_{a,m}|^2\delta(\omega_1-\omega^x_a)\delta(\omega_2-\omega^{xx}_m).
\end{equation}
According to Eq.~(\ref{Vk1}), the Coulomb matrix elements in the free carriers basis scales as $V_{ij,kl}\sim V^{-1} V_{{\bf k}_i{\bf k}_j,{\bf k}_k{\bf k}_l}\delta_{{\bf k}_i-{\bf k}_l,{\bf k}_j-{\bf k}_k}$. Then using Eq.~(\ref{Vxxx}), one finds that
\begin{eqnarray}\label{Kaux}
&~&\mathcal{K}(\omega_1,\omega_2)\sim V^{-2}\sum_{{\bf k}_i{\bf k}_j{\bf k}_k{\bf k}_l}|
			V^{eehe}_{{\bf k}_i,{\bf k}_j,{\bf k}_k, {\bf k}_i+{\bf k}_j+{\bf k}_k}|^2\;\;\;
\\\nonumber&~&\times
\delta(\omega_2-\omega^{e}_{{\bf k}_i} -\omega^{e}_{{\bf k}_j}-\omega^{h}_{{\bf k}_k}-\omega^{h}_{{\bf k}_l})
\\\nonumber&~&\times
\delta(\omega_1-\omega^{e}_{{\bf k}_i+{\bf k}_j-{\bf k}_k}-\omega^{h}_{{\bf k}_l})\sim V^2.
\end{eqnarray}
By taking into account that $\mathcal{K}$ should be normalized by the exciton and biexciton DOS and take square root, we introduce the prefactor $(V/v)^2$ into Eq.~(\ref{Veff}). The volume prefactor in Eq.~(\ref{Veff0}) can be obtained in the same way with the help of Eqs.~(\ref{Vxx}), (\ref{Vk1}), and (\ref{Vk2}).

\section{Representation of transition dipoles and interband Coulomb matrix elements in KW basis set}
\label{Appx_mtxl}

In this Appendix, we provide closed expressions for the exciton and biexciton transition dipoles and the interband Coulomb matrix elements used in the numerical calculations. To derive these expressions we follow the procedure outlined in Appendices~A-C of Ref.~[\onlinecite{piryatinski10}].

The second quantization is performed using the basis of KW states, $\{\Psi_i({\bf r})\}$ defined in Eq.~(\ref{KW_spinor}), by introducing the following field operators
\begin{eqnarray}\label{KW-sq}
	\hat\Psi({\bf r}) &=& \sum_i\left[\Theta(E_i)\Psi_i({\bf r}) c_i +\Theta(-E_i)\Psi_i({\bf r}) d^\dag_i\right],\;\;
\\\nonumber
	\hat\Psi^\dag({\bf r}) &=& \sum_i \left[\Theta(E_i)\Psi^*_i({\bf r}) c^\dag_i+\Theta(-E_i)\Psi^*_i({\bf r}) d_i\right],
\end{eqnarray}
where $\Theta(E)$ is the step function, $c_i$ and $d_i$ ($c^\dag_i$ and $d^\dag_i$) are conduction band electron and valence band hole annihilation (creation) operators, respectively.

Using this representation and the definition of the exciton and biexciton states given by Eqs.~(\ref{x-sq}) and (\ref{xx-sq}), it is straight forward to show that the transition matrix elements entering Eqs.~(\ref{nx}) and (\ref{nxx}) are
\begin{eqnarray}\label{mu_x}
	\mu^x_{a0} &=&\langle x_a|{\hat{\bf M}}|x_0\rangle = {\bf M}^{eh}_{qr} ,
\\\label{mu_xx}
	\mu^{xx}_{kl} &=&\langle xx_k|{\hat{\bf M}}|xx_l\rangle = [\delta_{rr'}\delta_{ss'}-\delta_{rs'}\delta_{r's}]
	\\\nonumber&\times&
			[{\bf M}^{ee}_{p'p}\delta_{qq'}-{\bf M}^{ee}_{q'p}\delta_{qp'}
					-{\bf M}^{ee}_{p'q}\delta_{pq'}+{\bf M}^{ee}_{q'q}\delta_{pp'}],
	\\\nonumber&+&[\delta_{pp'}\delta_{qq'}-\delta_{pq'}\delta_{p'q}]
	\\\nonumber&\times&
			[{\bf M}^{hh}_{s's}\delta_{rr'}-{\bf M}^{hh}_{s'r}\delta_{sr'}
					-{\bf M}^{hh}_{r's}\delta_{rs'}+{\bf M}^{hh}_{r'r}\delta_{ss'}],
\end{eqnarray}
respectively. In Eq.~(\ref{mu_x}), the exciton index $a=\{q,r\}$ and ${\bf M}^{eh}_{qr}$ is the interband electron-hole transition dipole matrix element calculated in the KW basis set. In Eq.~(\ref{mu_xx}), the biexciton indices are $k = \{p'q',r's'\}$ and $l = \{pq,rs\}$ as well as ${\bf M}^{ee}_{ij}$ and ${\bf M}^{hh}_{ij}$ are the matrix elements of the intraband transition dipole operator calculated using the electron and hole KW wave functions, respectively.

In the envelope function approximation, the KW transition matrix elements entering Eqs.~(\ref{mu_x}) and (\ref{mu_xx}) have the following generic form\cite{kang97}
\begin{align}
{\bf M}_{ij}&={\bf M}^{(1)}_{ij}+{\bf M}^{(2)}_{ij}, \nonumber \\
{\bf M}^{(1)}_{ij} &=\sum^4_{m=1}\int d{\bf r}\:[F^i_m({\bf r})]^*\hat{\bf p}F^j_m({\bf r}),
\nonumber \\
{\bf M}^{(2)}_{ij} &=P_l{\bf z}\int d{\bf r}\left (
 [F^i_1({\bf r})]^* F^j_3({\bf r})
+[F^i_3({\bf r})]^* F^j_1({\bf r})\right.\nonumber \\
&\left. -[F^i_2({\bf r})]^* F^j_4({\bf r})
-[F^i_4({\bf r})]^* F^j_2({\bf r})\right ),
\end{align}
where the indices $i,j$ can be associated with the electron and hole states, and ${\hat{\bf p}=-i\hbar {\boldsymbol \nabla}}$. $P_l$ stands for the longitudinal dipole moment component of the band-edge Bloch function, $u_m({\bf r})$, and ${\bf z}$ is the unit vector in $\langle 111\rangle$ direction of the PbSe (PbS) lattice. Both ${\bf M}^{(1)}$ and ${\bf M}^{(2)}$ can be evaluated analytically in the bulk limit, where ${\bf M}^{(1)}$ vanishes identically.\footnote{In the bulk limit the operator $\hat{\bf p}$ commutes with the bulk Hamiltonian, and therefore ${\bf M}^{(1)}$ becomes proportional to $\sum^4_{m=1}\int d{\bf r} \left [F^i_m({\bf r})\right]^* F^j_m({\bf r})$, which is zero identically due to the orthogonality of the initial and the final wave functions.}

Using the same approach as above, one can show that the Coulomb matrix elements entering Eqs.~(\ref{Lmbd}), (\ref{Veff}), (\ref{Veff0}), and (\ref{Kaux}) are
\begin{eqnarray}\label{Vxx}
V^{xx,0}_{l,0}&=&\langle xx_l|\hat{V}|x_0\rangle= V^{eehh}_{pqrs}- V^{eehh}_{pqsr},
\\\label{Vxxx}
V^{xx,x}_{k,a}&=&\langle xx_k|\hat{V}|x_a\rangle=
\left( V^{eehe}_{q'p'r'q}-V^{eehe}_{p'q'r'q}\right )\delta_{s',r}
\nonumber \\
&+&\left( V^{eeeh}_{q'p'q s'}-V^{eeeh}_{p'q'q s'}\right)\delta_{r',r}
\nonumber \\
&+&\left( V^{ehhh}_{q'r s'r'}-V^{ehhh}_{q'r r's'}\right) \delta_{p',q}
\nonumber \\
&+&\left( V^{ehhh}_{p'r r's'}-V^{ehhh}_{p'r r's'}\right) \delta_{q',q}.
\end{eqnarray}
These quantities depend on the long-range contributions whose matrix elements in the KW basis read
\begin{align}\label{CInt}
V_{ijkl}&=\int \int d{\bf r}_1 d{\bf r}_2\:\frac{e^2}{\epsilon|{\bf r}_1-{\bf r}_2|}
\sum^4_{m=1} \left[F^i_m({\bf r}_1)\right]^* F^l_m({\bf r}_1)
\nonumber \\
&\times\sum^4_{n=1} \left [F^j_n({\bf r}_2)\right]^* F^k_n({\bf r}_2).
\end{align}
Here, $\epsilon$ denotes the screened dielectric function values evaluated at the optical frequencies. For bulk PbSe and PbS, we set $\epsilon^{\rm bulk}$=23 and $\epsilon^{\rm bulk}$=17, respectively.\cite{kang97} The dielectric constant in the NCs has been evaluated using the following expression\cite{franceschetti99}
\begin{equation}
\epsilon^{NC}(d)=1+(\epsilon^{\rm bulk}-1)\frac{(E^{\rm bulk}_g+\Delta E)^2}{(E^{NC}_g(d)+\Delta E)^2},
\end{equation}
where $E^{\rm bulk}_g+\Delta E=2.73$~eV and 3.14~eV are the energy of the first pronounced absorption peak in the bulk PbSe and PbS, respectively.\cite{suzuki95,kanazawa98}

To derive the volume scaling of the Coulomb matrix elements given by Eq.~(\ref{CInt}), we assume the bulk limit in which the envelope functions become plane waves, i.e. $F^i_m({\bf r}_1)=F^i_m e^{i{\bf k}_i {\bf r}_1}/\sqrt{V}$. In this basis, the Coulomb matrix elements can be written as
\begin{eqnarray}
V_{ijkl}&=&\frac{e^2}{\epsilon V^2}\sum_{m,n}\left[F^i_m F^j_n\right]^* F^l_m F^k_n
\\\nonumber
&\times&\int\int d{\bf r}_1 d{\bf r}_2 \frac{1}{|{\bf r}_1-{\bf r}_2|}
e^{-i{\bf k}_i{\bf r}_1-i{\bf k}_j {\bf k}_2+i{\bf k}_k{\bf r}_2+i{\bf k}_l{\bf r}_1}.
\end{eqnarray}
The integral evaluation leads to the final expression
\begin{eqnarray}\label{Vk1}
V_{ijkl}&=&\frac{1}{V^2}V_{{\bf k}_i,{\bf k}_j,{\bf k}_k, {\bf k}_l}V\delta_{{\bf k}_i-{\bf k}_l,{\bf k}_j-{\bf k}_k},
\end{eqnarray}
where
\begin{eqnarray}\label{Vk2}
V_{{\bf k}_i,{\bf k}_j,{\bf k}_k, {\bf k}_l}=\frac{e^2}{\epsilon}\sum_{m,n}\left[F^i_m F^j_n\right]^* F^l_m F^k_n
\frac{4\pi}{|{\bf k}_i-{\bf k}_l|^2}.
\end{eqnarray}
Eqs.~(\ref{Vk1}) and (\ref{Vk2}) clearly demonstrate a general property that the Coulomb matrix elements used in the numerical calculations of the bulk limit scale inversely proportional to the volume.\cite{haug09}

The nonparabolicity in the employed $\bf{k}\cdot\bf{p}$-Hamiltonian is crucial for the accurate evaluation of the above Coulomb matrix elements. Specifically, summations over the spinor components in Eq.~(\ref{CInt}) implies that the interband Coulomb scattering amplitudes vanish exactly if the non-diagonal terms of the $\bf{k}\cdot\bf{p}$-Hamiltonian are set to zero, i.e., in strictly parabolic case.\cite{ridley87} In the bulk, where quasimomentum $k$ is a ``good'' quantum number, diagonal and off-diagonal matrix elements of the $\bf{k}\cdot\bf{p}$-Hamiltonian scale as $k^2$ and $k$, respectively. At high energies, where the diagonal elements dominate over the off-diagonal ones, the latter can be treated perturbatively giving $k^{-1}\sim \omega^{-1/2}$ as the contribution of the hole (electron) states to a high energy electron (hole) wave function.


\end{document}